\documentclass[
 reprint,
superscriptaddress,
groupedaddress,
unsortedaddress,
 amsmath,amssymb,
 aps,
 showkeys,
prb,
]{revtex4}

\usepackage{amsmath,amssymb}
\usepackage{color}
\usepackage{graphicx}% Include figure files
\usepackage{dcolumn}% Align table columns on decimal point
\usepackage{bm}% bold math
\usepackage{hyperref} % Required for hyperlinks
\usepackage{wasysym} % Required to insert math
\usepackage{epsfig,subfigure,float,pseudocode,multirow,footmisc,rotating}
\usepackage[nottoc, notlof, notlot]{tocbibind}
\usepackage{mathtools}
\usepackage{multirow}

\definecolor{grey}{rgb}{0.5,0.5,0.5}

\begin{document}

\preprint{APS/123-QED}

%%%% Article title to be placed here
\title{Automated Analysis of Behavioural Variability and Filial Imprinting of Chicks\\ (\textit{G. gallus}) using Autonomous Robots}

%\author{A. Gribovskiy}
%%\email{alexey.gribovskiy@epfl.ch}
%\affiliation{LSRO, EPFL, Lausanne, Switzerland}
%
%\author{F. Mondada}
%%\email{Francesco.Mondada@epfl.ch}
%\affiliation{LSRO, EPFL, Lausanne, Switzerland}
%
%\author{JL. Deneubourg}
%%\email{jldeneub@ulb.ac.be}
%\affiliation{USE, ULB, Bruxelles, Belgium}
%
%\author{L. Cazenille}
%%\email{leo.cazenille@univ-paris-diderot.fr}
%\affiliation{Univ Paris Diderot, Sorbonne Paris Cit\'e, LIED, UMR 8236, 75013, Paris, France} 
%\affiliation{Sorbonne Universit\'es, UPMC Univ Paris 06, CNRS, ISIR, F-75005 Paris, France}
%
%\author{N. Bredeche}
%%\email{nicolas.bredeche@upmc.fr}
%\affiliation{Sorbonne Universit\'es, UPMC Univ Paris 06, CNRS, ISIR, F-75005 Paris, France}
%
%\author{J. Halloy}
%%\email{jose.halloy@univ-paris-diderot.fr}
%\affiliation{Univ Paris Diderot, Sorbonne Paris Cit\'e, LIED, UMR 8236, 75013, Paris, France} 

\author{%%%% Author details
A. Gribovskiy$^{1}$,
F. Mondada$^{1}$,\\
JL. Deneubourg$^{2}$,
L. Cazenille$^{3,4}$,\\
N. Bredeche$^{4}$,
%J. Halloy$^{3}$ jose.halloy@univ-paris-diderot.fr}
J. Halloy$^{3}$}
\email{jose.halloy@univ-paris-diderot.fr}

%%%%%%%%% Insert author address here
\address{$^{1}$ LSRO, EPFL, Lausanne, Switzerland\\
$^{2}$ USE, ULB, Bruxelles, Belgium\\
$^{3}$ Univ Paris Diderot, Sorbonne Paris Cit\'e, LIED, UMR 8236, F-75205 Paris, France\\
$^{4}$ Sorbonne Universit\'es, UPMC Univ Paris 06, CNRS, Institute of Intelligent Systems and Robotics (ISIR), F-75005 Paris, France}
%$^{4}$ Sorbonne Universit\'es, UPMC Univ Paris 06, UMR 7222, ISIR, F-75005, Paris, France\\
%$^{5}$ CNRS, UMR 7222, ISIR, F-75005, Paris, France}

%%%% Insert corresponding author and its email address}
%\corres{J. Halloy\\
%\email{jose.halloy@univ-paris-diderot.fr}}

%%%% Abstract text to be placed here %%%%%%%%%%%%
\begin{abstract}
\scriptsize
Inter-individual variability has various impacts in animal social behaviour. This implies that not only collective behaviours have to be studied but also the behavioural variability of each member composing the groups. To understand those effects on group behaviour, we develop a quantitative methodology based on automated ethograms and autonomous robots to study the inter-individual variability among social animals. We choose chicks of \textit{Gallus gallus domesticus} as a classic social animal model system for their suitability in laboratory and controlled experimentation. Moreover, even domesticated chicken present social structures implying forms or leadership and filial imprinting. We develop an imprinting methodology on autonomous robots to study individual and social behaviour of free moving animals. This allows to quantify the behaviours of large number of animals. We develop an automated experimental methodology that allows to make relatively fast controlled experiments and efficient data analysis. Our analysis are based on high-throughput data allowing a fine quantification of individual behavioural traits. We quantify the efficiency of various state-of-the-art algorithms to automate data analysis and produce automated ethograms. We show that the use of robots allows to provide controlled and quantified stimuli to the animals in absence of human intervention. We quantify the individual behaviour of 205 chicks obtained from hatching after synchronized fecundation. Our results show a high variability of individual behaviours and of imprinting quality and success. Three classes of chicks are observed: imprinted with variability, indifferent to the imprinting stimuli (i.e the robots) and chicks that avoid the imprinting stimuli. Our study shows that the concomitant use of autonomous robots and automated ethograms allows detailed and quantitative analysis of behavioural patterns of animals in controlled laboratory experiments.
\end{abstract}
%%%%%%%%%%%%%%%%%%%%%%%%%%%

%%%% Keyword entries to be placed here %%%%
\keywords{Filial imprinting, \textit{Gallus gallus domesticus}, High-Throughput Ethology, Autonomous Robots, Collective Behaviour}

%%%%%%%%%% Insert the texts which can accomdate on firstpage in the tag "fmtext" %%%%%

%\begin{fmtext}
%\section{Insert A head here}
%%%%% Insert A head here
%
%This demo file is intended to serve as a ``starter file''
%for rsproca journal papers produced under \LaTeX\ using
%rsproca.cls v1.5e.
%
%\subsection{Insert B head here}
%%%%% Insert B head here
%Subsection text here.
%
%\subsubsection{Insert C head here}
%%%%% Insert C head here
%Subsubsection text here.
%
%\section{Equations}
%
%Sample equations.
%
%%%% Numbered equation
%\begin{align}\label{1.1}
%\begin{split}
%\frac{\partial u(t,x)}{\partial t} &= Au(t,x) \left(1-\frac{u(t,x)}{K}\right)-B\frac{u(t-\tau,x) w(t,x)}{1+Eu(t-\tau,x)},\\
%\frac{\partial w(t,x)}{\partial t} &=\delta \frac{\partial^2w(t,x)}{\partial x^2}-Cw(t,x)+D\frac{u(t-\tau,x)w(t,x)}{1+Eu(t-\tau,x)},
%\end{split}
%\end{align}
%
%\begin{align}\label{1.2}
%\begin{split}
%\frac{dU}{dt} &=\alpha U(t)(\gamma -U(t))-\frac{U(t-\tau)W(t)}{1+U(t-\tau)},\\
%\frac{dW}{dt} &=-W(t)+\beta\frac{U(t-\tau)W(t)}{1+U(t-\tau)}.
%\end{split}
%\end{align}
%
%%%%% Unnumbered equation
%\begin{eqnarray}
%\frac{\partial(F_1,F_2)}{\partial(c,\omega)}_{(c_0,\omega_0)} = \left|
%\begin{array}{ll}
%\frac{\partial F_1}{\partial c} &\frac{\partial F_1}{\partial \omega} \\\noalign{\vskip3pt}
%\frac{\partial F_2}{\partial c}&\frac{\partial F_2}{\partial \omega}
%\end{array}\right|_{(c_0,\omega_0)}\notag\\
%=-4c_0q\omega_0 -4c_0\omega_0p^2 =-4c_0\omega_0(q+p^2)>0.
%\end{eqnarray}
%\end{fmtext}

%%%%%%%%%%%%%%% End of first page %%%%%%%%%%%%%%%%%%%%%

\maketitle

\twocolumngrid
\section{Introduction}
Many species of animal live in group and are capable of making decisions while maintaining group coherence \cite{sumpter2010collective,camazine2003self}. Animal living in groups are constrained by what they can perceive thus most of the time they cannot have a global view of the group and their environment. Yet those species are capable of remarkable group behaviours. The question arises as to understand what kind of mechanism allows the group to perform so well given their individual limitations. Several forms of social structures exists that go from all individuals apparently having the same weight in the group to various forms of leadership and social hierarchy.
\\

Among the many group-living species, the chicken (\textit{Gallus gallus domesticus}) is an interesting animal model: the chickens are social animals and chicks present a social attachment to their siblings and to their mother.
Indeed, in precocial birds such as chickens, the formation of social attachments occurs in the first days after hatching. This process is used as a model for the study of learning \cite{horn1985memory} because the chicks' environment and conditions prior to hatching can be controlled and standardized. Hence, the period immediately after hatching is likely to be particularly useful for studying social motivation and learning.
\\

This social structure allows to address the question of the role of leadership in group decision-making. Chicks are attracted to each other and at the same time they tend to remain in a relatively close vicinity to their mother that they can also follow as a group. The question arises as to what makes the group coherent and what is the importance of the leader, the hen, in group decision making and group coherence.
The development of filial behaviour can be separated in terms of two interacting processes: filial motivation and filial imprinting. Filial motivation causes chicks to approach and follow conspicuous objects with certain general properties like colours, shapes, sizes, and movement patterns.
In a natural environment, chicks are attracted toward the hen and the other chicks of the group.
Filial imprinting is a learning process through which the chick comes to restrict this behaviour to focus to particular stimuli \cite{Bateson1966,sluckin1970imprinting,bolhuis1991mechanisms,van1996framework}. As chicks become progressively more familiar with an object, they tend to approach this object more \cite{horn1985memory}. Thus, filial motivation determines the propensity to make social attachments, while imprinting establishes preference.
\\

In ethological studies one of the long-standing interests is to understand social communication, relationships and structures.
Until recently, to study these mechanisms researchers had used simple specially designed mock-ups whose behaviour can be controlled so as to trigger a response of the animals.
But nowadays availability of low-cost miniaturized computer chips, motors and sensors allowed these artificial models to become sophisticated and reliable robotic devices that can be used to test hypotheses \cite{knight2005animal}.
For instance, robots were used to study male territorial instinct in dart-poison frogs \cite{narins2005cross} and mate selection in Tungara frogs \cite{taylor2008faux}, to test ideas about nest mate recognition in brush turkeys \cite{goth2004social}, and the predator avoidance by ground squirrels \cite{rundus2007ground}.
A number of recent works in ethology have successfully used robots to investigate individual and collective animal behaviours, in particular by creating mixed robot-animals societies: robots were mixed with chicks \cite{Gribovskiy2010}, cockroaches \cite{sempo2006integration,halloy2007social}, fruit flies \cite{zabala2012simple}, killifish \cite{phamduy2014fish} and zebrafish \cite{polverino2013zebrafish}.
\\

\begin{figure}[]
\centering\includegraphics[width=8.30cm]{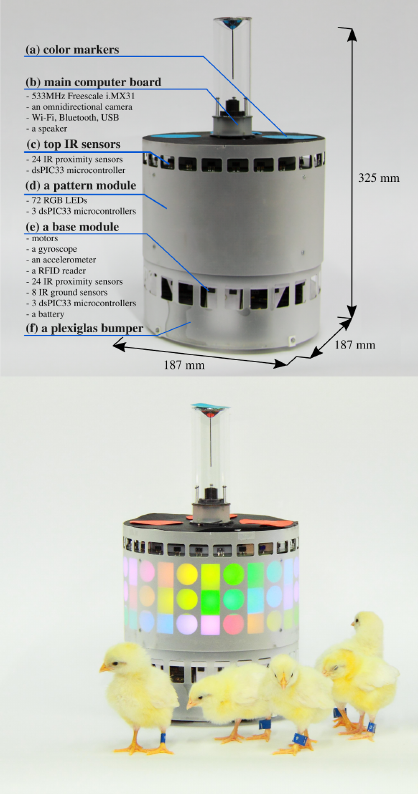}
\caption{{\bf The PoulBot autonomous robot.} The robot developed to test individual and social behaviour among chicks is based on the modular research robot \textit{marXbot} \cite{Gribovskiy2010}. This robot is equipped with a plastic shell that prevent chicks from being roll over or stuck in the tracks. The upper side of the shell allows to display a predefined colour pattern. The tube on the top is a $360^{\circ}$ omnidirectional camera that allows the robot to perceive visually its surrounding. The small rectangular holes allow infra-red sensors to detect objects or animals in close vicinity.}
\label{Fig1}
\end{figure}

\begin{figure}[h]
\centering\includegraphics[width=17.35cm]{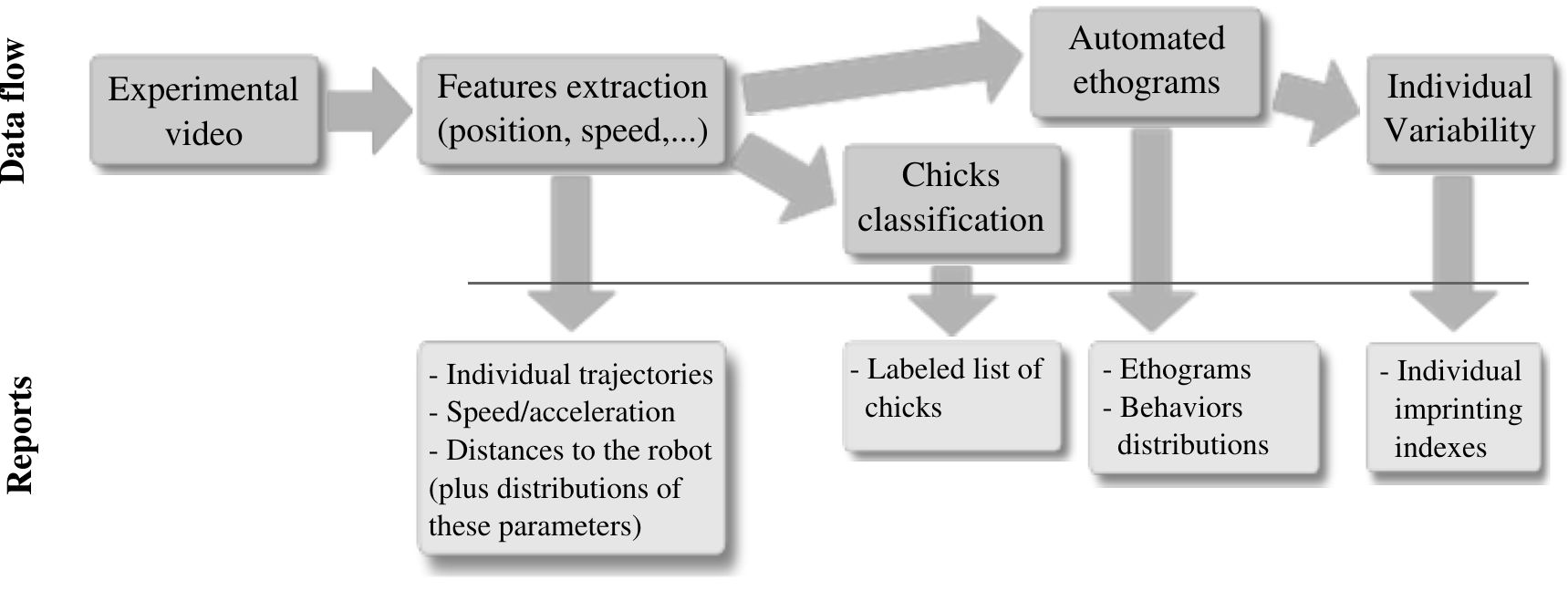}
\caption{{\bf Data analysis process} First we analyse the experimental videos and extract a number of relevant features (position, speed, acceleration, distances) at every time-step. Behavioural analysis is based on these features. Our analysis is performed in two steps: on one hand, we classify the chicks according to the success of the filial imprinting on the robot, and on the other hand, we automatically generate individual ethograms and behavioural sequences. Lastly, we use the generated ethograms to compare and classify chicks by their following behaviour.}
\label{Fig2}
\end{figure}

Here, we address these questions by using a robot (cf. Fig.~\ref{Fig1}) that will become a surrogate mother to individual chicks. To build a social attachment between the chicks and the robot we use the filial imprinting mechanism. After hatching a critical period exists where the chick will learn to build social attachment to specific objects. Classical ethology has shown that this special form of learning works for any object remaining close to the chicks and presenting specific colour, movement and sound patterns.
In this study, we develop an imprinting methodology involving autonomous robots to study individual behaviour of free moving animals. Thanks to this methodology we are able to quantify the individual behaviours and social responses to the robot.
We also develop automated experimental methodologies that allow to make fast and efficient data analysis, and compare to existing ones \cite{branson2009high,kabra2012jaaba}.
A workflow of this methodology is found in Fig.~\ref{Fig2}.
The analysis is based on high throughput data allowing fine quantification of individual behavioural traits. Thanks to this method we can make quantitative assessment of individual variability of a large number of individuals.
\\

We investigate whether the imprinting quality of chicks can be influenced by social bonding. After the imprinting process, the chicks are randomly put in groups of 6 individuals for 20 days. We then perform a new quantification of individual behaviour traits, and compare them to original results. Additional experiments include experiments during the period of time the chicks are put into groups, where entire groups of chicks interact with a robot, but the associated results are not investigated in this article.
Here, we quantify the individual characteristics for 205 animals. The results show a large variability of behavioural patterns. One of the main variability is related to the success of imprinting. Three types of chicks are observed: imprinted, indifferent and avoiders. Among the successfully imprinted chicks we still observe a large variability of behavioural patterns.
Finally we show that the use of autonomous robots and automated ethograms allow detailed quantitative studies of animal behavioural patterns as it has also been shown with Drosophila flies and bees \cite{zabala2012simple,landgraf2011analysis}.

\clearpage
\onecolumngrid

\twocolumngrid

%%%%%%%%%%%%%%%%%%%%%%%%%%%%%%%%%%%%%%%%%%%%%%%%%%%%%%%%%%%%%%%%%%%%%%%%%%%%%%%%%%%%%%%%%%%%%%%%%%%%%%%%%
%%%%%%%%%%%%%%%%%%%%%%%%%%%%%%%%%%%%%%%%%%%%%%%%%%%%%%%%%%%%%%%%%%%%%%%%%%%%%%%%%%%%%%%%%%%%%%%%%%%%%%%%%
%%%%%%%%%%%%%%%%%%%%%%%%%%%%%%%%%%%%%%%%%%%%%%%%%%%%%%%%%%%%%%%%%%%%%%%%%%%%%%%%%%%%%%%%%%%%%%%%%%%%%%%%%
%%%%%%%%%%%%%%%%%%%%%%%%%%%%%%%%%%%%%%%%%%%%%%%% RESULTS %%%%%%%%%%%%%%%%%%%%%%%%%%%%%%%%%%%%%%%%%%%%%%%%
%%%%%%%%%%%%%%%%%%%%%%%%%%%%%%%%%%%%%%%%%%%%%%%%%%%%%%%%%%%%%%%%%%%%%%%%%%%%%%%%%%%%%%%%%%%%%%%%%%%%%%%%%
%%%%%%%%%%%%%%%%%%%%%%%%%%%%%%%%%%%%%%%%%%%%%%%%%%%%%%%%%%%%%%%%%%%%%%%%%%%%%%%%%%%%%%%%%%%%%%%%%%%%%%%%%
%%%%%%%%%%%%%%%%%%%%%%%%%%%%%%%%%%%%%%%%%%%%%%%%%%%%%%%%%%%%%%%%%%%%%%%%%%%%%%%%%%%%%%%%%%%%%%%%%%%%%%%%%

\section{Results}

\subsection{Filial imprinting as a means of social bonding with robots} \label{sec:classificationRes}
The first step in the behavioural analysis is to classify the chicks according to the success of the filial imprinting on the robot.
Filial imprinting corresponds to the propensity of the chicks to be attracted by the robot and to follow it as if it was the hen mother.
We observe two different kind of non-imprinted chicks: indifferent chicks that ignore the presence and movement of the robot, and avoiders that run along the walls, possibly running away from the robot.
Imprinting success is assessed by looking at the behavioural response of the chicks towards the robot.
\\

\begin{center}
\begin{figure}[h]
\centering\includegraphics[width=14.00cm]{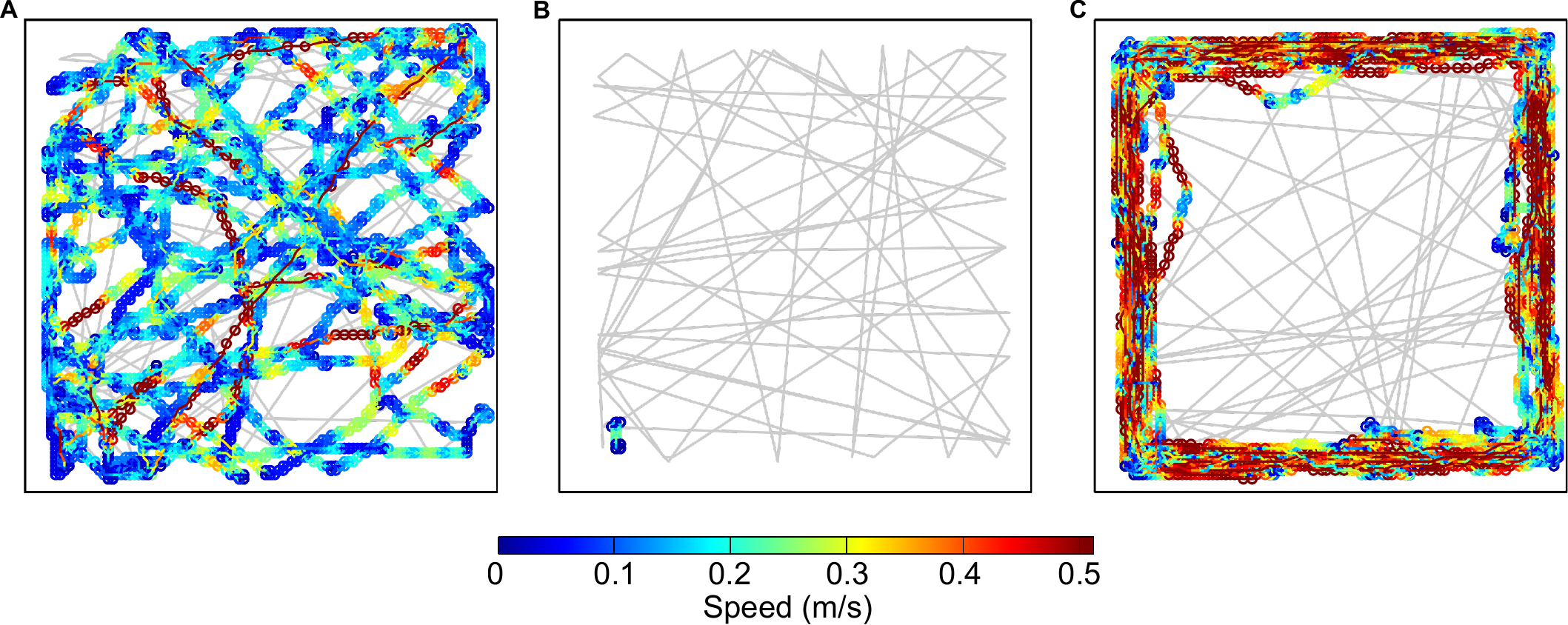}
\caption{{\bf Speeds of three different chicks depending on spatial position} Panels A, B, C show the speed of chicks depending on their spatial position in the arena, in three different settings, presenting different behavioural types. Panel A corresponds to a chick that has been classified as imprinted. Panel B corresponds to a chick that does not move and is considered as non imprinted. Panel C corresponds to a chick that avoids the robot and runs along the walls of the setup. The grey lines represent the robot trajectories in the arena that does not take into account the chick presence.}
\label{Fig3}
\end{figure}
\end{center}

\begin{figure}[h]
\centering\includegraphics[width=8.00cm]{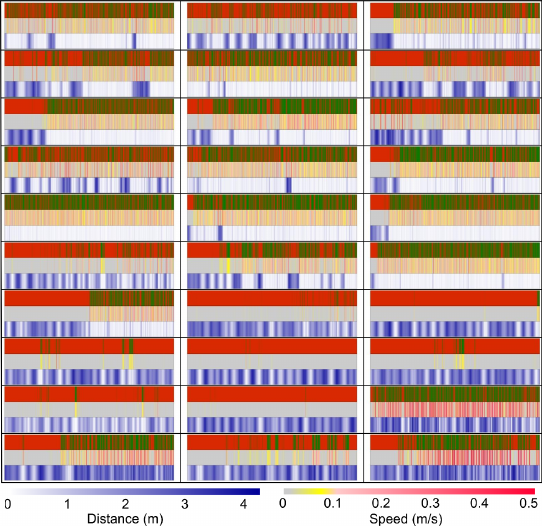}
\caption{{\bf Distribution of chicks and visually sorted behavioural patterns.} For each individual, the first line is the distribution of stops (in red) and walking behaviour (in green), the second line is the distribution of speed ($m s^-1$), and the last line is the distribution of the distance ($m$) between the robot and the individual. Only 30 individuals are considered. All experiments have a duration of 1 hour. The bars are drawn in the experimental chronological order from top left to right.}
\label{Fig4}
\end{figure}

\begin{figure}[h]
\centering\includegraphics[width=8.00cm]{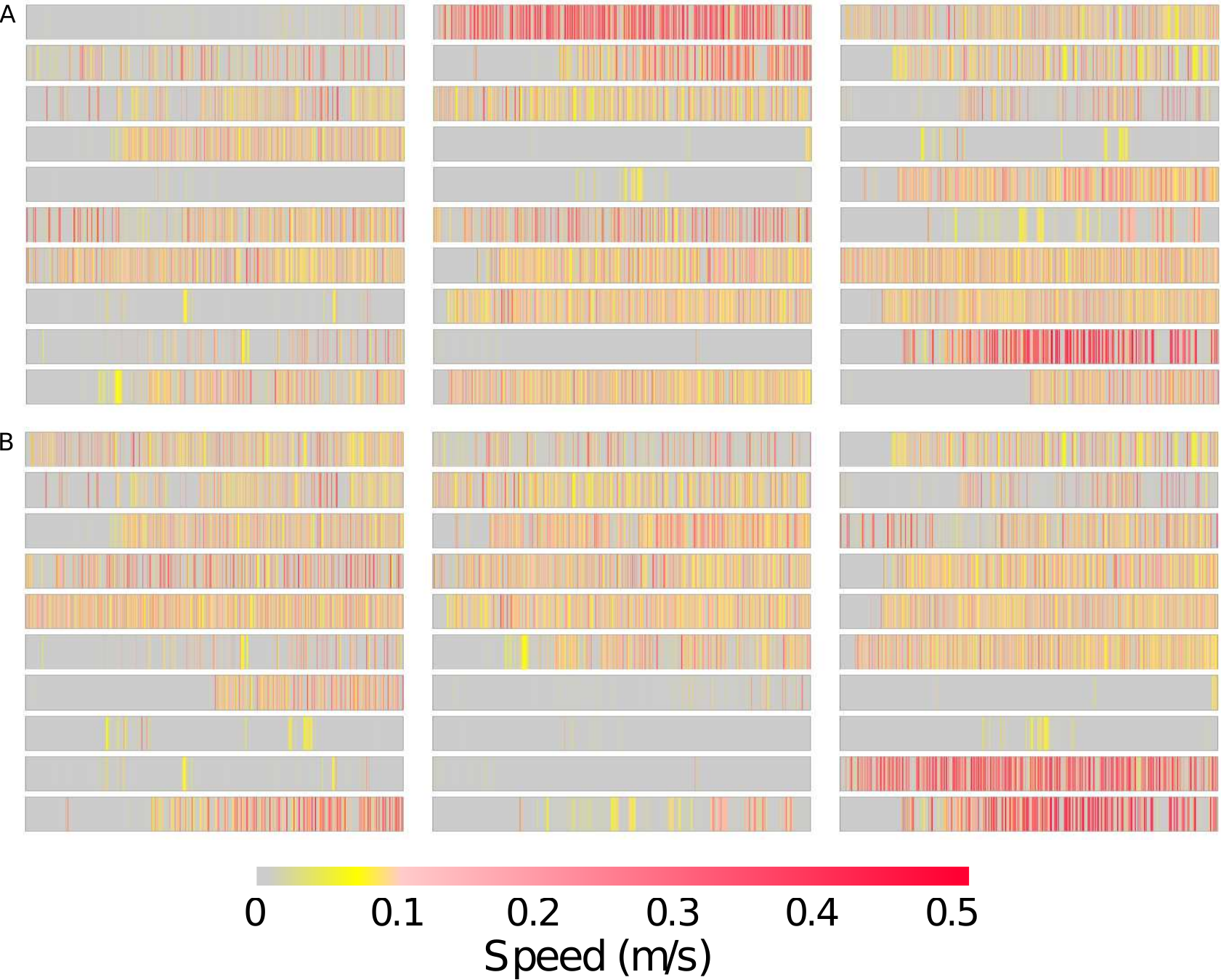}
\caption{{\bf Speed distribution of chicks and visual classification.}  Panel A shows the visualization of speed distributions for 30 individuals during tests. The bars are drawn in the experimental chronological order from top left to right. Panel B shows that we can easily visually classify the animals into three categories. First,  those that move with variable speed. Second, those that nearly do not move. Third, those that move at high speed nearly all the time. We will see below that these three categories are consistent with quantitative statistical clustering methods.}
\label{Fig5}
\end{figure}

Figure \ref{Fig3} shows both the speed of three different chicks depending of their spatial position in the arena, and the robot trajectories. The figure shows the three typical behaviour responses of a chick to the robot: a chick can be imprinted (Panel A), indifferent (Panel B), or avoider (Panel C).
The distributions of chicks speed, their distance toward the robot, and their walking behaviour (whether they are walking or stopping), for 30 individuals, is shown in Fig.~\ref{Fig4}. Imprinted and avoider chicks tend to alternate a walking behaviour with small stops. Indifferent chicks stay in place in the arena, with very few apparent walking behaviour. Imprinted chicks tend to be close to the robot. Indifferent and avoider chicks tend to be distant from the robot.
The three typical behaviour responses of a chick to the robot correspond to different distributions of chick speeds, as shown in Fig.~\ref{Fig5}: imprinted chicks move with variable speed, indifferent chicks do not move much, avoider chicks tend to move at higher speed than imprinted chicks.
\\

The classification of the behavioural response of each chick toward the robot can be done by a human observer looking at all the videos.
However, this represents a tedious and long qualitative work that would greatly benefit from automation. Only $31.22\%$ of all experiments (64 individuals) is classified by a human observer. The rest of the experiments is classified automatically, using an algorithm to classify the data automatically extracted from the videos by image analysis.
\\

\begin{figure}[]
\centering\includegraphics[width=7.0cm]{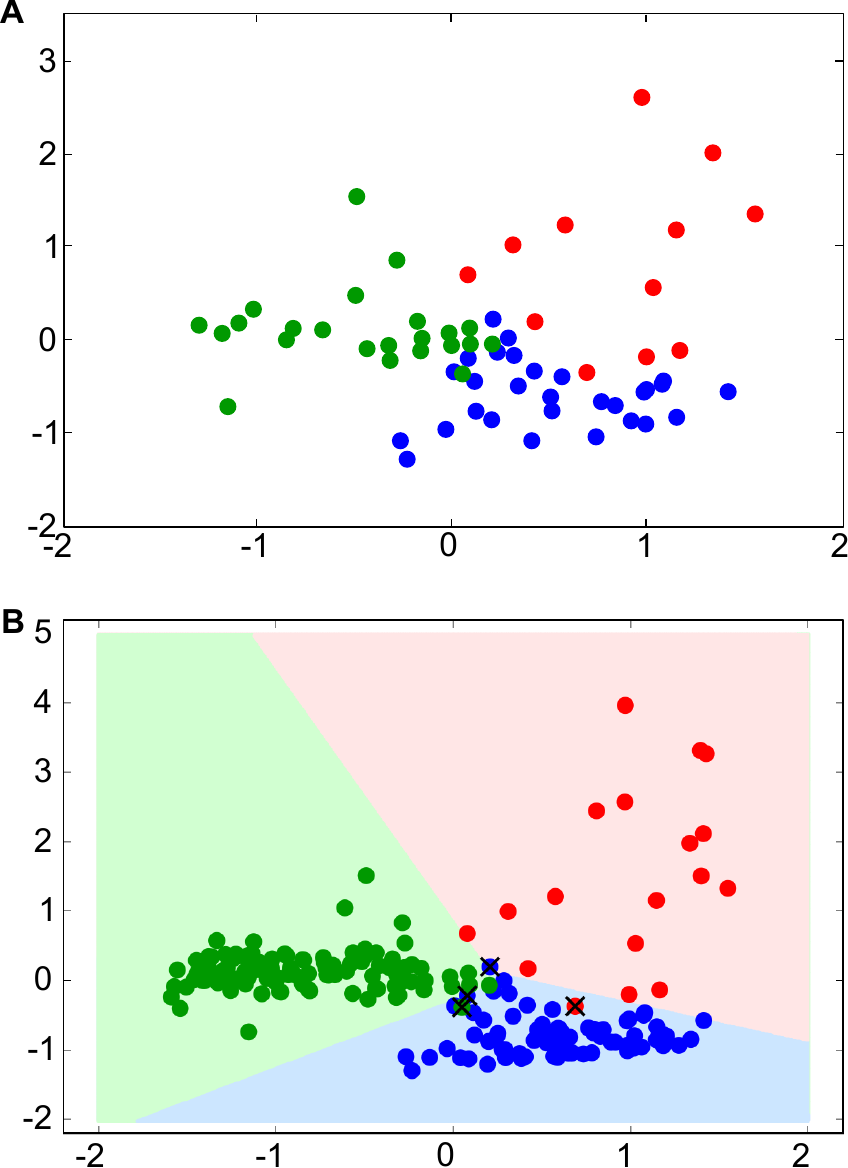}
\caption{{\bf Classification of the chicks according to the filial imprinting success on the robot.}
The classification process used the following features: \textit{mean distance between a robot and a chick}, \textit{mean speed of a chick}, and its \textit{standard deviation}.
We reduced the dimensionality of the related feature vector from three to two by using Principal Components Analysis (PCA) \cite{Bishop2006}. Panels A and B show the values of the feature vector for the considered experiments.
Panel A shows the training dataset for the Support Vector Machine (SVM) classifier ($31.22\%$ of all experiments: 64 individuals, priorly classed by human observation).
The green points corresponds to chicks that are attracted and follow the robot (imprinted), the red points to chicks that avoid the robot (avoider) and the blue to chicks that do no react to the presence of the robot (indifferent).
Panel B shows how the trained classifier divides the plane into three regions and shows the result of the classification on the whole data set (205 individuals). The green area corresponds imprinted chicks, the pink area to avoider chicks and the blue to chicks indifferent chicks. Crossed points (4 individuals) correspond to misclassification after checking the whole dataset by human inspection. Misclassification happens when the chicks show a mixed behavioural type that is also difficult to interpret by a human observer. }
\label{Fig6}
\end{figure}

Because it is possible for a human observer to classify with good accuracy the type of chick, we select a supervised learning method. We use a linear support vector machine (SVM) classifier, an algorithm often used to solve problems in classification, regression, and novelty detection \cite{Vapnik1995,Meyer2003,Bishop2006}. This method is first trained on the human-classified dataset, separating experiments into three classes: imprinted, avoider, indifferent.
The classification process used the following features: \textit{mean distance between a robot and a chick}, \textit{mean speed of a chick}, and its \textit{standard deviation}.
The results of the classification are presented in Fig.~\ref{Fig6}. Panel A shows the training dataset composed of $31.22\%$ of the all data (64 individuals). The green points correspond to the imprinted chicks, the blue points to the indifferent chicks, and red points to the avoider chicks. The experiments have a duration of one hour, and were repeated for $205$ different chicks.
Panel B  of Fig.~\ref{Fig6} shows the results of the trained classifier on the whole data set (205 individuals). The classifier present an accuracy of 98.05\% (201/205). The crossed points corresponds to misclassification (4 individuals). These misclassification are the consequence of mixed behavioural pattern. Even for human observers, these individuals are difficult to categorize because, during the test, they present characteristics of imprinting or avoider behavioural traits.
\\

As a result of individual tests we observed $55.12\%$ (113/205) imprinted chicks, $36.10\%$ (74/205) indifferent and $8.78\%$ (18/205) of avoider chicks.

%\clearpage
\onecolumngrid
\twocolumngrid

\clearpage

\subsection{Automated Ethograms}\label{sec:automatedEthograms}
Ethograms are typically used to describe animal behaviour. They are made usually by human experts consistently documenting behaviour from observations. The two main approaches are focal sampling (in which the human expert make systematic observation of the behaviour of an individual -- it is a precise but time-expensive method) and scan sampling (in which the observer only uses observation recorded at regular intervals to classify behaviours -- it is less time-consuming compared to focal sampling, but also less accurate).
Ethograms done by human observation are time consuming and tedious if one wants to obtain quantitative results. We present a combined methodology that performs detailed quantitative studies of single individual repeated for a large number of animals (205 individuals).
\\

Recently, a number of ethological studies have used similar methodologies to automatically generate ethograms of animal behaviour from behavioural data \cite{anderson2014toward,branson2009high,kabra2012jaaba,gerencser2013identification,de2012computerized}. This process of automatic behaviours detection is call "Automated ethology", and produces "automated ethograms".
It is usually done in two steps.
The first step is called "Segmentation": the trajectories are split into segments. Each segment is labelled independently from the others, using either classification or clustering algorithms.
In the second step, a label (or "class") is assigned to each segment, describing which kind of behaviour the chick is exhibiting as the corresponding time. Two kind of algorithms can be used for labelling: classification algorithms and clustering algorithms. For each segment, these algorithms use a set of features (statistical measures) of the observed individual to find a corresponding label. Before the labelling process, classification algorithms need to be trained with a training set of segments already labelled by a human. This training set is used as reference (or \textit{a priori} knowledge) to guide the labelling process. Clustering algorithm do not need to be trained beforehand, and only uses the similarity of the segments' features to group them together into clusters. Each cluster correspond to a different label.
\\

The segmentation strategy can impact the reliability of the labelling process of trajectories compared to a labelling done by humans. This is especially the case when segments incorporate periods of time when the chicks exhibit several different behaviour. On the other hand, when segments are too small, it can add noise into the features used for labelling. In an optimal segmentation, segments end when a chick changes its behaviour.
A typical segmentation method used in the literature is the $N$-steps window segmentation: trajectories containing $M$ time-steps are split into $M/N$ segments each with a size of $N$ time-steps. An extreme case is the $1$-step window segmentation, where every time-step is a new segment. Note that the choice of $N$ can influence the reliability of a segmentation: if $N$ is too big, segments can include period of times where the chicks exhibit several different behaviours. On the other hand, $N$-step window segmentation with a low $N$ can be more sensitive to noise.
\\

\begin{figure}[]
\centering\includegraphics[width=8.3cm]{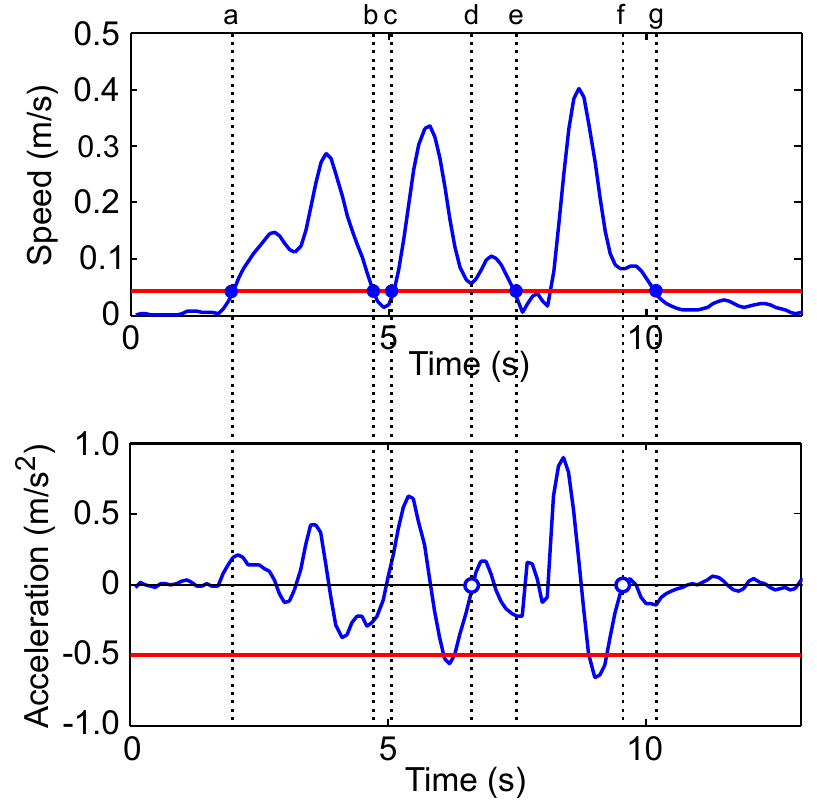}
\caption{{\bf Defining how trajectories are segmented.} The criteria for segmentation is based on speed and acceleration. Two thresholds (red lines) are defined for speed and for acceleration. The speed threshold is used to detect stops. Below the threshold ($44 mm.s^{-1}$) the animal is considered as not moving. To further segment intervals during which the animal is moving, the criteria is based on the acceleration. First we define a deceleration threshold ($-500 mm.s^-2$) below which we consider that the animal is going to change type of movement. After being below the threshold, when the acceleration equals to 0, the trajectory is segmented. Then a new segment is started where the animal has changed its movement type. In the shown example of experimental data, doted line (a) marks the transition from stop to movement, (b) from movement to stop, (c) from stop to movement, (d) change of movement type, (e) from movement to stop, (f) change of movement type and (g) from movement to stop.}
\label{Fig7}
\end{figure}

We introduce a more robust segmentation method, the 'Threshold crossing' method.
We make two hypothesises, supported by human observations: first, that chicks reduce their speed below a given threshold when they change behaviour; second, that chicks accelerate when they begin to exhibit a given behaviour, and decelerate when they change behaviour. This allow us to introduce an original segmentation method, that we call 'Threshold crossing'. This segmentation method is based on the speed and acceleration of a chick (Fig.~\ref{Fig7}): time-steps when the speed or acceleration of a chicks are below predefined thresholds are considered as stops.
\\

The segments obtained are then labelled using state-of-the-art algorithms of the literature, either using classification algorithms: Decision Trees (DT), Decision Trees with Random Forest (DT/Forest), k-Nearest Neighbours (k-NN), Support Vector Machines (SVM); or a clustering algorithm: K-Means. These methods are described in detail in Sec. \ref{sec:behaviouralMeasurments} and \ref{sec:classificationAndClusteringOfTrajectories}.
For each segment, the classification and clustering algorithms use the following features (statistical measures) to find a label (corresponding to the behaviour of the chick): the mean speed of the chick, the acceleration of the chick, the distance between the chick and the robot, the distance between the chick and the wall of the arena, and the distance traveled by the chick during the experiment.
\\

Based on human observation, we separated chick behaviours into seven classes:
\begin{description}
	\item[stops] the chick stops moving
	\item[joins] the chick, initially distant from the robot, joins the robot (with a higher speed than when following it)
	\item[follows] the chick is close to the robot and follows it
	\item[in front] the chick goes in front of the robot
	\item[loops] the chick goes around the robot and make a full loop around it
	\item[bumped] the robot pushes gently the chick out of its way
	\item[other] the chick has a behaviour that is not related to the robot
\end{description}
Note that the robot is programmed to constantly move, even if a chick is on its trajectory.
This set of classes correspond to the labels used by the classification algorithms during the labelling process.
\\

We validated the automated labelling of chicks trajectories by computing the error compared to human-labeled trajectories: the collection of human-labeled behavioural sequences was split into two parts: the first part was used during the training step (learning base), and the second part was used to perform this validation. As the observed error of classification is very low ($0.04$), we can say that the automated labelling is very close to the labelling done by humans.
\\

% TODO
\begin{table*}[]
\caption{{Table of mean Absolute percentage error (MAPE) of different combinations of segmentation and classification methods, compared to human-labelled ethograms.Three segmentation methods are considered: (1) the \textbf{threshold crossing} described in Sec. \ref{sec:behaviouralMeasurments} and Fig.~\ref{Fig7}, (2) the \textbf{$N$-step windows} where segments have a fixed size of $N$ time-steps. The classification and clustering algorithms tested are described in Section \ref{sec:automatedEthograms} : \textbf{DT} stands for \textit{Decision Trees}, \textbf{kNN} for \textit{k-Nearest Neighbours}, \textbf{SVM} for \textit{Support Vector Machines}. The MAPE values are computed using the formula: $MAPE = \frac{1}{n} \sum_{t=1}^n \lvert \frac{A_t - F_t}{A_t} \rvert $ with $A_t$ the human-labelled reference values and $F_t$ the values obtained by classification methods. The best-performing scores are put in bold. }}
\label{table1}

	\scriptsize
	\centering

	\begin{tabular}{l l l | l | l | l | l | l | l |}
		\cline{4-9}
		& & & \multicolumn{6}{c|}{\textbf{Segmentation Method}} \\

		\cline{4-9}
		& & & \multicolumn{2}{|c|}{Threshold crossing} & \multicolumn{2}{|c|}{1-step window} & \multicolumn{2}{|c|}{5-steps window} \\

		\cline{4-9}
		& & & MAPE & std.dev. & MAPE & std.dev. & MAPE & std.dev. \\

		\cline{1-9}

		%\multicolumn{1}{|c|}{\multirow{4}{*}{\parbox[t]{2cm}{\textbf{Algorithm}}}}
		\multicolumn{1}{|c|}{\multirow{5}{*}{\parbox[t]{0.4cm}{\rotatebox[origin=c]{90}{\textbf{Algorithm}}}}}
			& \multicolumn{1}{|c|}{\multirow{4}{*}{\parbox[t]{4cm}{\textbf{Classification (supervised)}}}}
		& DT \cite{Breiman1984} & $0.20$ & $0.04$ & $0.19$ & $0.04$ & $2.11$ & $0.91$ \\
		\cline{3-9}
		\multicolumn{1}{|c|}{}
			& \multicolumn{1}{|c|}{}
		& DT/Forest \cite{Breiman2001} & $\mathbf{0.04}$ & $0.00$ & $\mathbf{0.04}$ & $0.01$ & $0.89$ & $0.12$ \\
		\cline{3-9}
		\multicolumn{1}{|c|}{}
			& \multicolumn{1}{|c|}{}
		% NOTE : cite is a review... Is that okay ?
		& kNN \cite{kantardzic2011data} & $5.40$ & $1.60$ & $5.56$ & $2.87$ & $6.37$ & $4.22$ \\
		\cline{3-9}
		\multicolumn{1}{|c|}{}
			& \multicolumn{1}{|c|}{}
		% NOTE : correct cite ?
		& SVM \cite{Bishop2006} & $0.12$ & $0.02$ & $0.11$ & $0.11$ & $1.48$ & $0.45$ \\

		\cline{2-9}
		\multicolumn{1}{|c|}{}
			& \multicolumn{1}{|c|}{\multirow{1}{*}{\parbox[t]{4cm}{\textbf{Clustering (unsupervised)}}}}
		% NOTE : cite is a review... Is that okay ?
		& K-Means \cite{kantardzic2011data} & $8.57$ & $3.84$ & $12.99$ & $4.76$ & $12.73$ & $4.98$ \\

		\hline
	\end{tabular}
\end{table*}

We measured the reliability of the different methods of segmentation and labelling presented compared to human-labeled trajectories. In addition to the 'Threshold crossing' segmentation method, we also tested the $N$-step window segmentation for several values of $N$: $N=1$ and $N=5$. These results are compiled in Table.~\ref{table1} using Mean Absolute Percentage Error (MAPE) values. The MAPE values are computed using the formula: $MAPE = \frac{1}{n} \sum_{t=1}^n \lvert \frac{A_t - F_t}{A_t} \rvert $ with $A_t$ the human-labelled reference values and $F_t$ the values obtained by classification methods.
The best-performing classification method is the Decision Trees with Random Forests method, when using a Threshold crossing segmentation or a 1-window segmentation. Methods using the $5$-step window segmentation method have a smaller reliability than methods using the $1$-step window segmentation: the segments obtained by $5$-step window segmentation can incorporate periods of time when the chick exhibit several different behaviours. It should also be noted that chicks behavioural traces often include shorts stops of the chicks when they change behaviour. These stops can have a duration inferior to $5$ time steps, and may impact negatively the performance of the $5$-step window segmentation.
The K-Means clustering algorithm give promising results, but is outperformed by the tested classification algorithms. It is not surprising, as this is a clustering algorithm: it is not trained beforehand, and cannot use any \textit{a-priori} information (except the number of labels).
\\

\begin{figure*}[]
\centering\includegraphics[width=14.00cm]{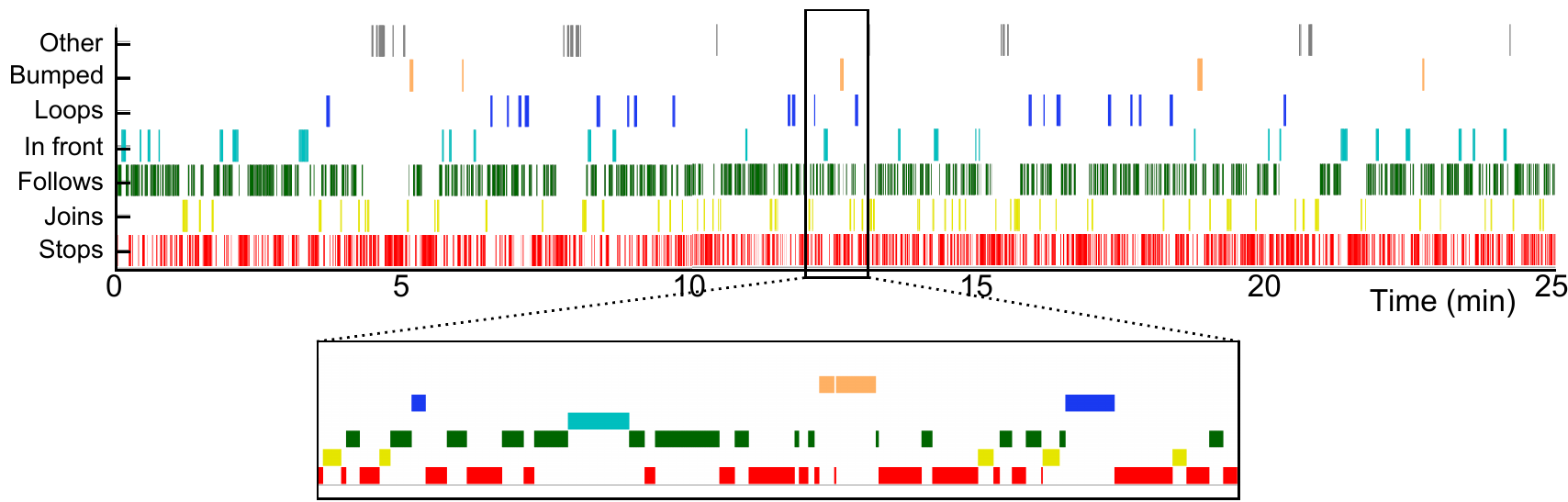}
\caption{{\bf Automated ethogram of an individual chick.} The method of behavioural classification based on Decision Trees with Random Forests allows producing automated ethograms. The figure shows, for a test of 25 minutes, the sequence of behaviours performed by the animal while the robot is constantly moving in a straight line at a constant speed of $70$~mm/s and reflecting randomly on the walls of the set-up. The enlargement shows a behavioural sequence of 1 min. The behaviours are classified in seven categories namely: stop (red), joins (yellow), follow (green), in front(light blue), loops (dark blue), bumped (orange) and other (grey). "Stop" corresponds to  moments when the chick is considered as not moving (see method), "joins" corresponds to a chick that is further away from the robot and that joins it, "follows" correspond to an animal that is close to the robot and follows it, "in front" is when the chick goes in front of the robot, "loop" is when the chicks makes a loop around the robot, bumped is when the robot pushes the chick out of its way and "other" means anything else than the previous categories.}
\label{Fig8}
\end{figure*}

Figure \ref{Fig8} shows an automated ethogram of an individual chick using the best-performing segmentation and classification methods: "Threshold crossing" and Decision Trees with Random Forest.
\\

\begin{figure*}[]
\centering\includegraphics[width=14.00cm]{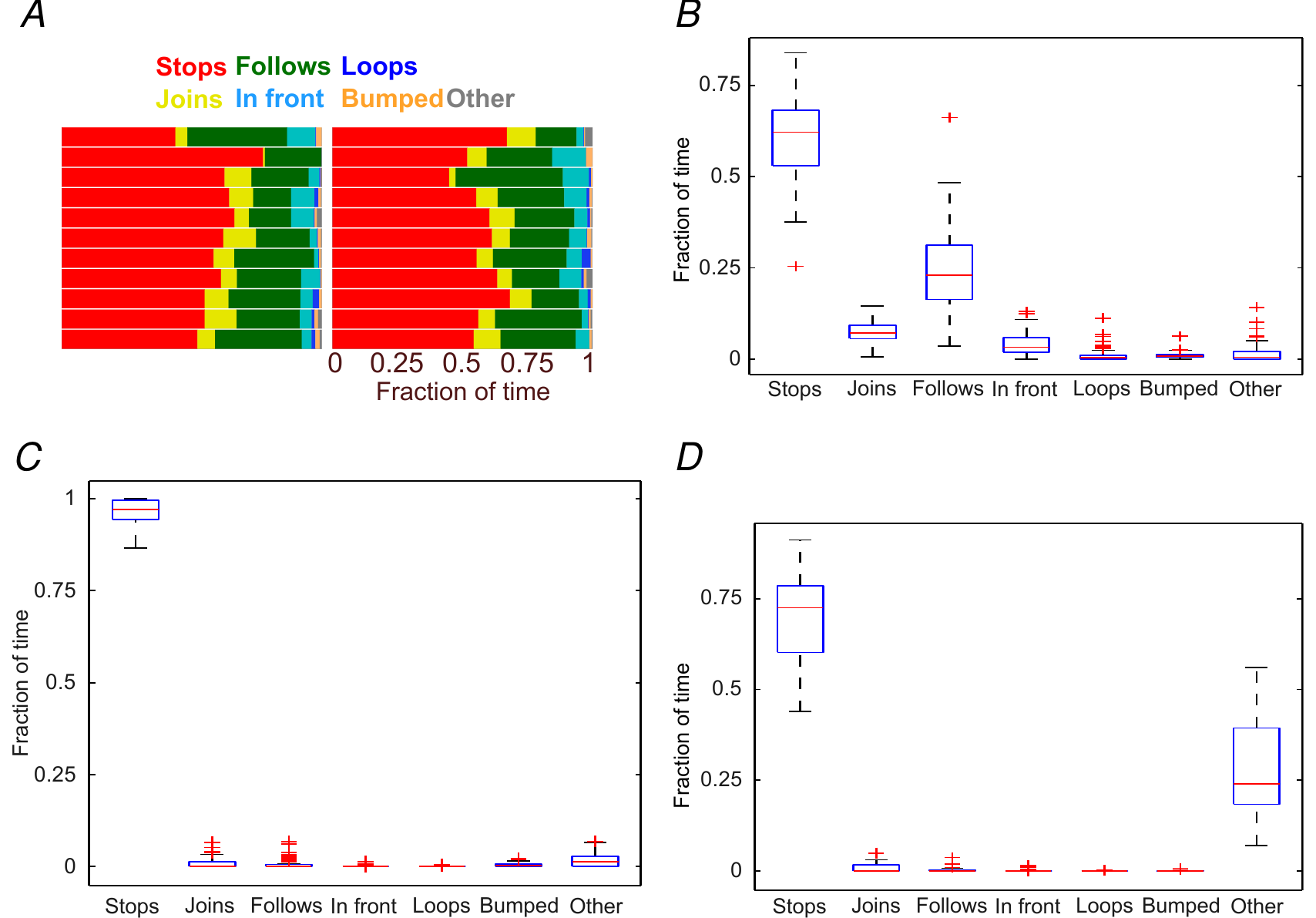}
\caption{{\bf Behaviours distribution for all chicks} for chicks classified as imprinted (Panel B, 113 individuals), for chicks classified as indifferent (Panel C, 74 individuals), and for chicks classified as avoider (Panel D, 18 individuals).
  Panel A: Examples of behavioural time budget of imprinted chicks computed from the sequence ethograms. The fractions of time that chicks spend in each type of behaviour are represent as stacked histograms. The colour code correspond to each type of behaviour. Each bar corresponds to a 25 min test of different individuals. The most important integrated fraction is the time chicks spend not moving although they are following the robot.}
\label{Fig9}
\end{figure*}

Figure \ref{Fig9} shows the distribution of durations of each behaviour for all imprinted chicks (113 individuals), all indifferent chicks (74 individuals) and all avoider chicks (18 individuals), over three series of experiments. Statistical analysis (Kruskal-Wallis one-way analysis of variance \cite{kruskal1952use}) shows that the three datasets are significantly different (p=$ 2.2375\times10^{-5}$).
\\

\begin{figure}[]
\centering\includegraphics[width=8.3cm]{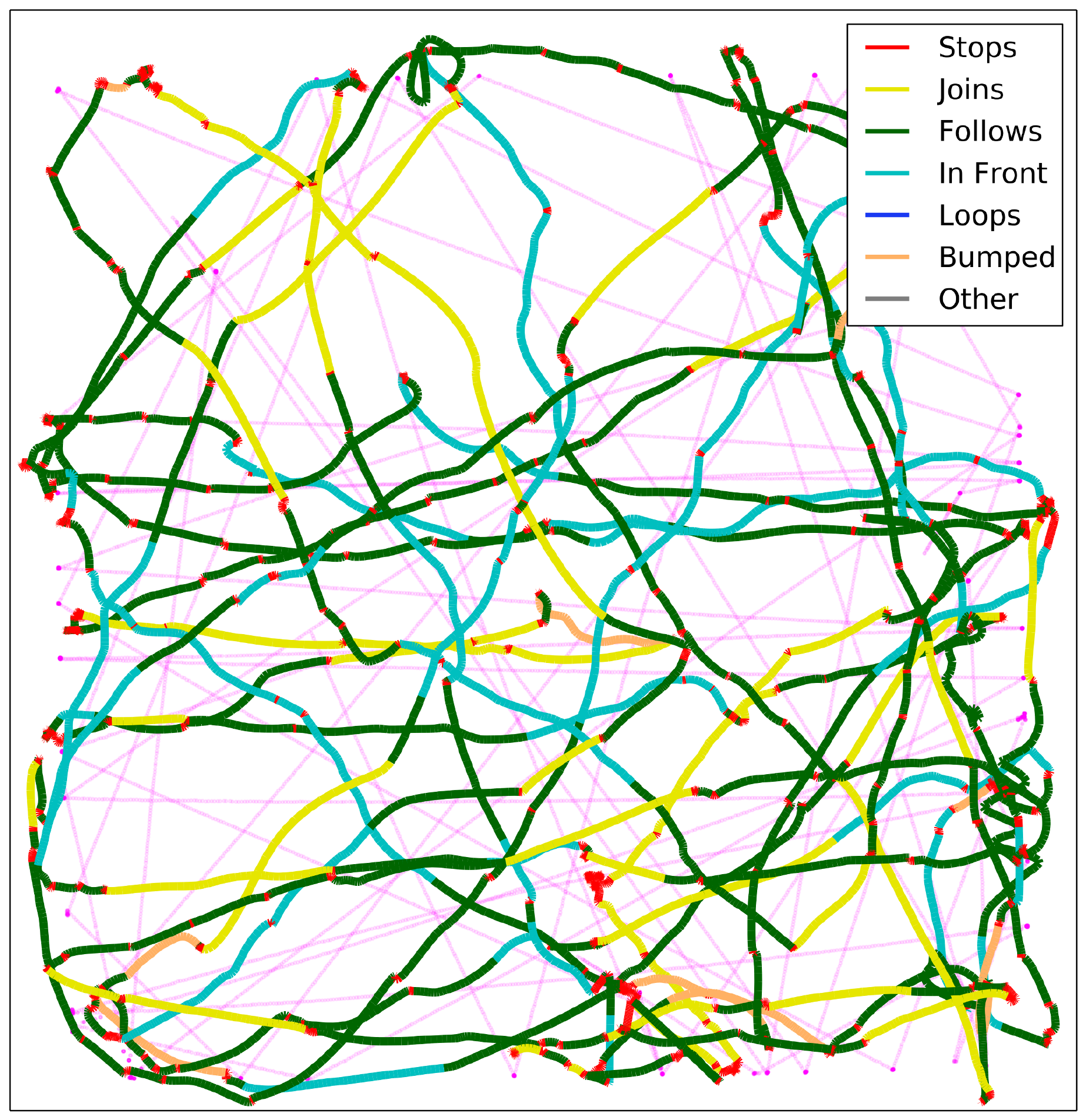}
\caption{{\bf Example of trajectories of one imprinted chick and the robot in the experimental arena.} The magenta lines represent the robot trajectories in the arena, while the coloured lines represent the trajectory of an imprinted chick. Labelling of the chick trajectories by colours was done automatically using Decision Trees with each colour corresponding to specific observed behaviour.}
\label{Fig10}
\end{figure}

Figure \ref{Fig10} presents an example of a trajectory of an imprinted chick. This trajectory is segmented and coloured according to the observed behaviour of the chick. The chick tends to stop when it is changing direction, or when it is changing it's behaviour.
\\

\begin{figure}[]
\centering\includegraphics[width=7.0cm]{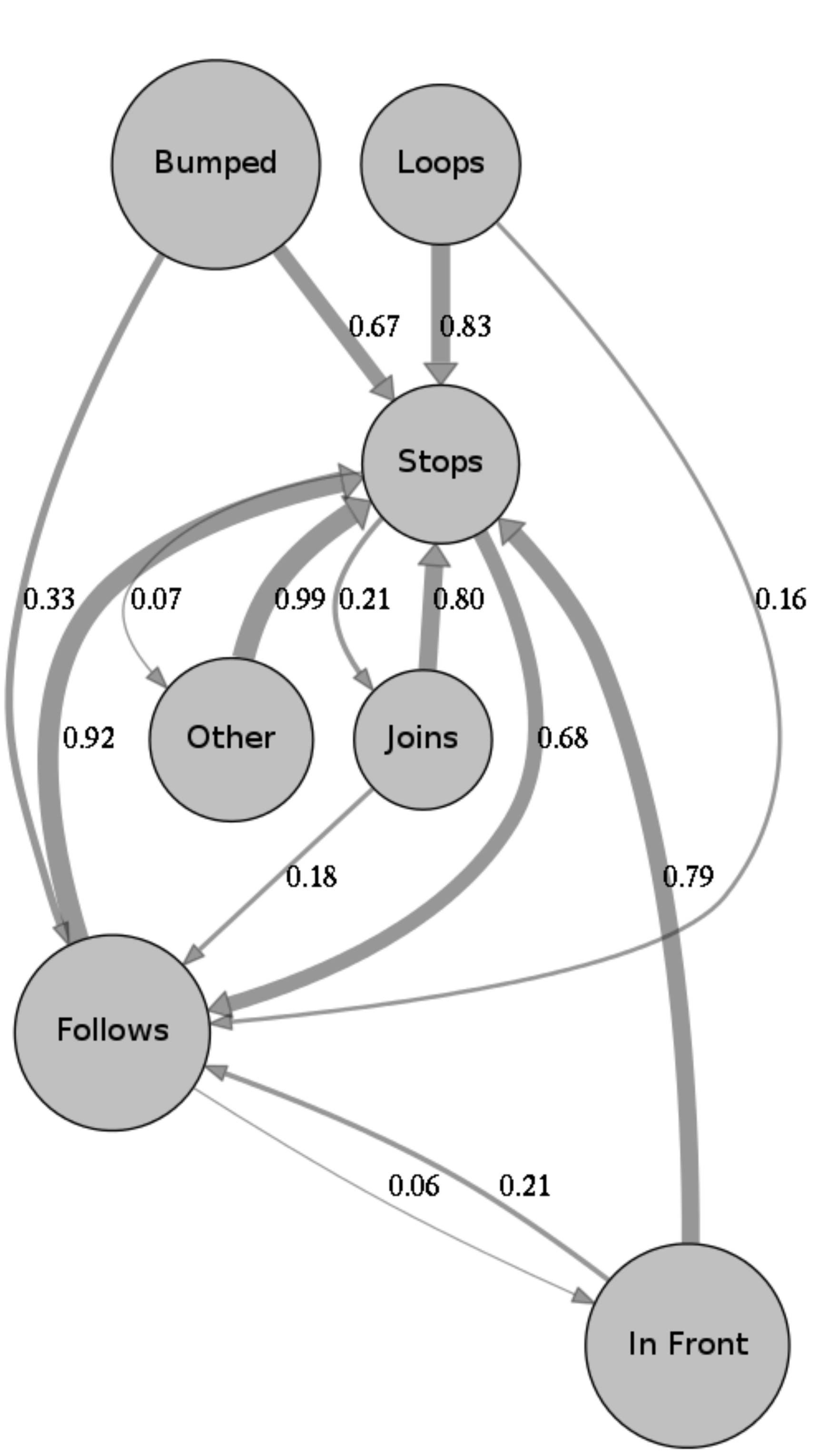}
\caption{{\bf Sequence of behaviours according to the transition matrix probabilities.} The possible sequences of behaviours are shown according to the transition matrix (see Table~\ref{table2}). The thickness of a transition arrow is proportional to its occurrence probability. Only transitions with an occurrence probability higher than 0.05 are represented. Self-transitions are not taken into account.}
\label{Fig11}
\end{figure}

Automated ethograms generated by our methodology can take two forms.
Individual ethograms, in the form of behavioural sequences (Fig.~\ref{Fig8}), describe the evolution of the behaviour of a single chick (in one given experiment) with respect to time.
The global dynamics of chicks behaviour for all imprinted chicks can be described using the transition frequencies between the seven behavioural patterns considered (Table~\ref{table2}).
Transition frequencies can be used to build global ethograms, in the form of Finite State Machines, or state transition diagrams, as shown in Fig.~\ref{Fig11} (self-transitions are not taken into account).
According to these figures, chicks tend to stop when they are changing their behaviour.

\begin{table*}[!ht]
\caption{{\bf Transition (from rows to columns) frequencies among behavioural patterns for imprinted chicks (113 individuals).}  Probabilities were computed from a total of 66247 events. Probability values higher than $0.05$ are put in bold.}
\scriptsize
\begin{tabular}{|l| *{7}{m{1.3cm}|} }
\hline\textbf{Behaviours} & \textit{Stops}  & \textit{Joins} & \textit{Follows} & \textit{In front} & \textit{Loops} & \textit{Is bumped} & \textit{Other} \\
\hline\textit{Stops}  &     & $\mathbf{0.213}$ & $\mathbf{0.677}$ & $0.021$ &$0.001$ &$0.018$ &$\mathbf{0.07}$ \\
\hline\textit{Joins}  & $\mathbf{0.795}$ &   & $\mathbf{0.182}$ & $0.014$ &$0.006$ &$0.002$ &$0.001$ \\
\hline\textit{Follows} & $\mathbf{0.922}$ & $0.000$ &  & $\mathbf{0.057}$ &$0.018$ &$0.003$ &$0.000$ \\
\hline\textit{In front} & $\mathbf{0.787}$ & $0.002$ & $\mathbf{0.210}$ &  &$0.000$  &$0.001$ &$0.000$ \\
\hline\textit{Loops}  & $\mathbf{0.829}$ & $0.004$ & $\mathbf{0.163}$ & $0.002$ &  &$0.002$ &$0.000$ \\
\hline\textit{Is bumped} & $\mathbf{0.666}$ & $0.002$ & $\mathbf{0.328}$ & $0.002$ &$0.002$ &  &$0.000$ \\
\hline\textit{Other}  & $\mathbf{0.995}$ & $0.005$ & $0.000$ & $0.000$ &$0.000$  &$0.000$  & \\
\hline
\end{tabular}
\label{table2}
\end{table*}

\subsection{Quantification of individual variability} \label{sec:indivVar}
In this section, we show that is it possible to quantify the imprinting quality of the chicks toward the robot, by using the individual ethograms described in Sec. \ref{sec:automatedEthograms}.
Only data from imprinted and indifferent chicks are considered, as avoiders chicks avoid any imprinting stimuli toward the robot: the imprinting process fails and the imprinting quality is null.
\\

\begin{figure}[]
\centering\includegraphics[width=8.3cm]{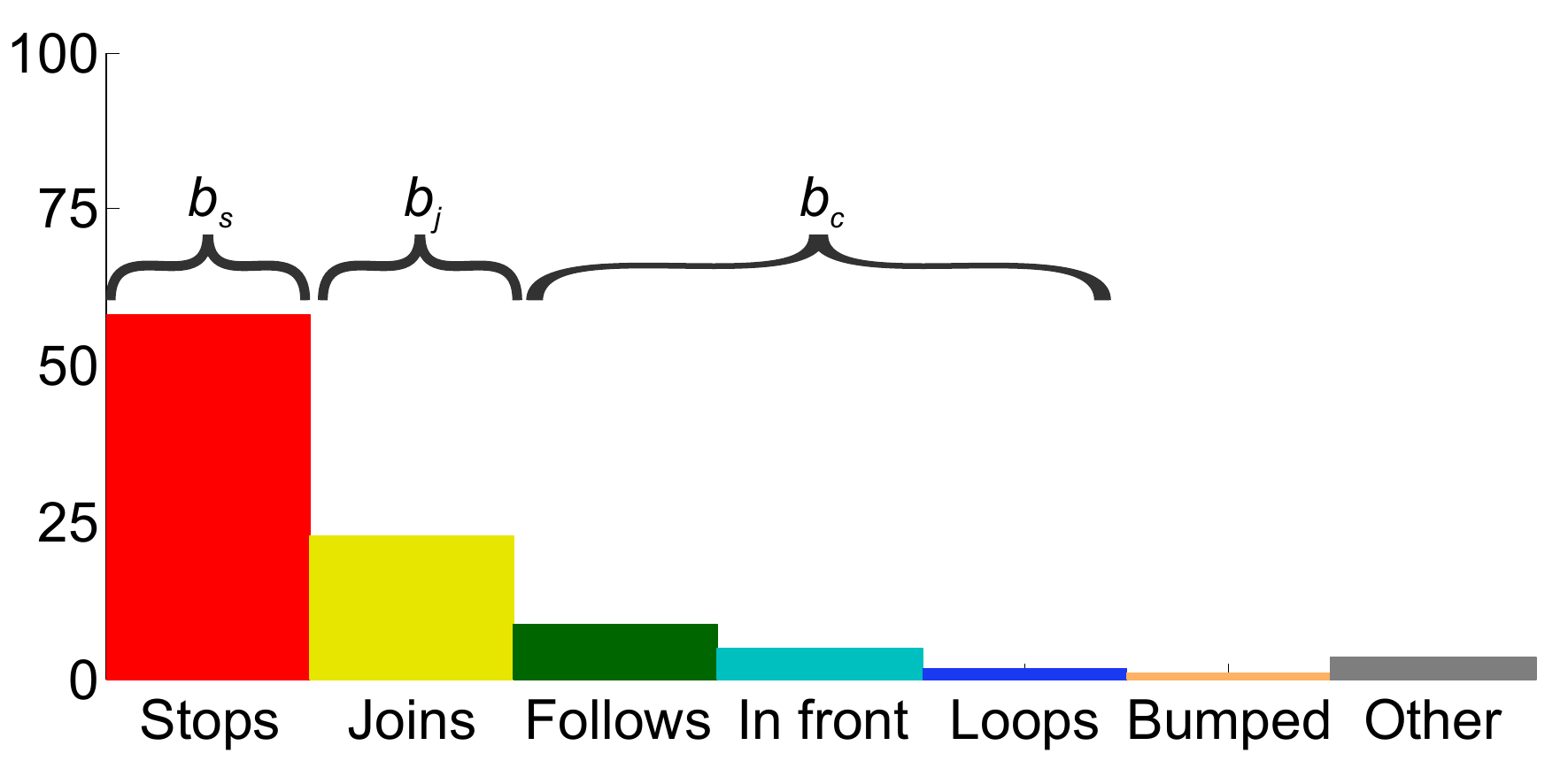}
\caption{{\bf Construction of a quantitative imprinting index based on the behavioural time budget.} The histograms represent the amount of time spend by a chick in each type of behaviour. For each individual, it is possible to build a vector, $(b_s,b_j,b_c)$, taking into account the most important behaviours. A PCA analysis show that $97.26\%$  of the variance in the data is explained by the first principal component. Then, we build an imprinting index on the base of a linear transformation of the vector. This imprinting index allows to qualitatively sort each individual on an imprinting scale. Only data from imprinted and indifferent chicks are used.}
\label{Fig12}
\end{figure}

The global distribution of behaviours for every chick can be found in Fig.~\ref{Fig12}.
Our objective is to define an imprinting index, corresponding to the measure of individual attachment of a chick toward the robot. We use information from the distribution of behaviour of each chicks to compute this index. As we want to define a one-dimensional index, this information needs to be compressed.
\\

We introduce a new representation of the behaviour of each chicks, similar to the one used in \ref{Fig12} and previous sections, but with a reduced dimensionality ($3$ instead of $7$): for each chick, a vector of three values $(b_s,b_j,b_c)$ represents respectively the portion of time the chick (1) rests, (2) runs toward the robot and (3) stays close to the robot. These values are presented in Fig.~\ref{Fig12}. In this representation, the "Bumped" behaviour is ignored because it only occurs rarely. The "Other" behaviour is ignored because it does not correspond to a clearly identified behaviour.
\\

To obtain a representation of chick behaviour with a further reduced dimensionality,
we performed principal component analysis (PCA) on the dataset composed of $(b_s,b_j, b_c)$ vectors for all chicks: $97.26\%$ of the variance in this dataset is explained by the first principal component. This shows that the first principal component is sufficient to quantify the behaviour of a chick.
The first principal component is computed as follow: $b_sp_s+b_jp_j+b_cp_c$, where $p_s=0.7510$,$p_j=-0.0975$ and $p_c=-0.6531$.
We define an imprinting index, corresponding to the measure of individual attachment of a chick toward the robot, by a linear transformation of the first principal component:
$$ii = ii_0 -(b_sp_s+b_jp_j+b_cp_c),$$
where $ii_0$ is chosen to separate imprinted and non-imprinted (indifferent and avoiders) sets.
\\

This imprinting index can be used to sort all chicks by the ``imprinting force'' of a chick toward the robot.
\\

\begin{figure}[]
\centering\includegraphics[width=8.3cm]{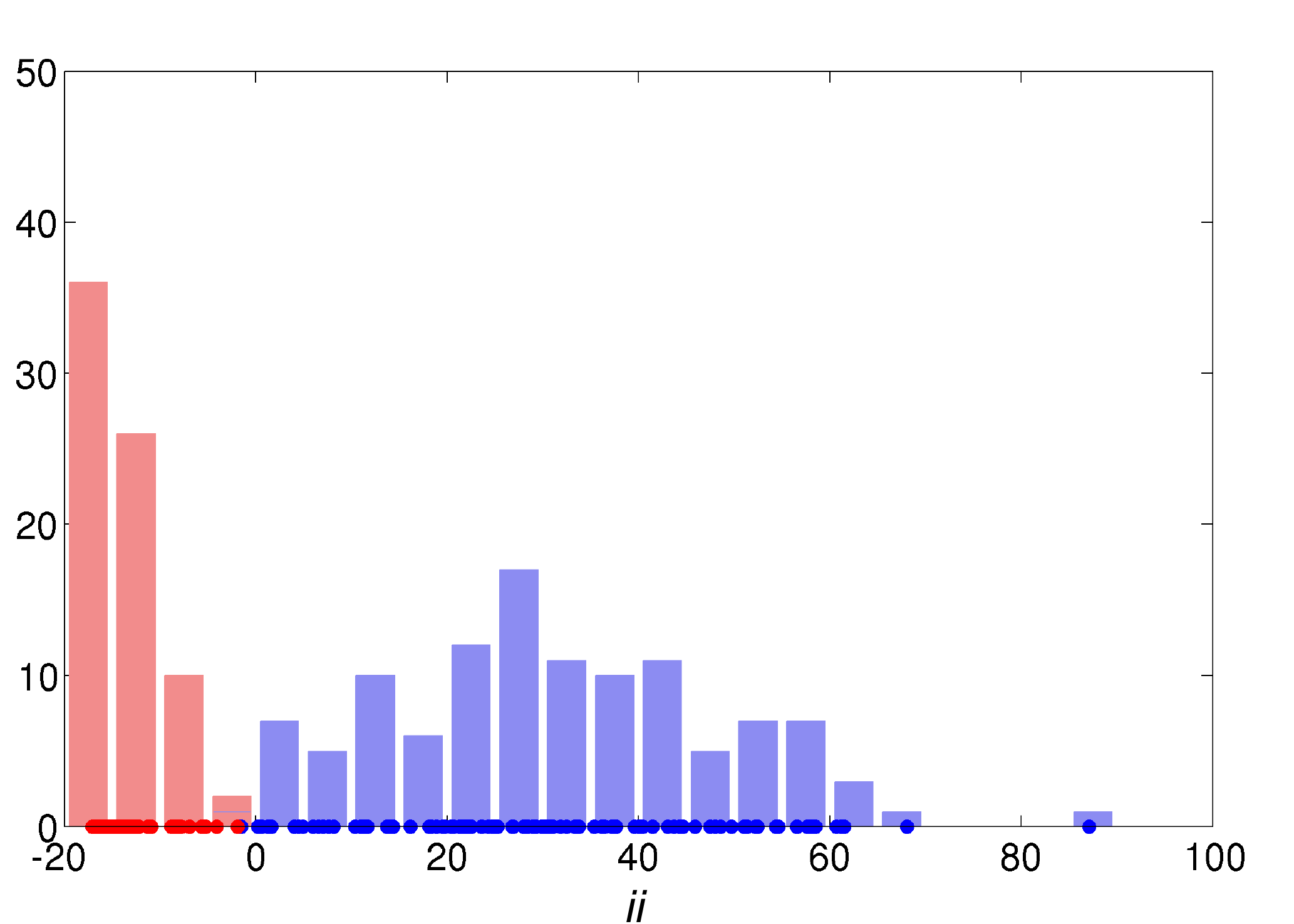}
\caption{{\bf Imprinting indexes} Values and distribution of imprinting indexes for all imprinted and indifferent chicks (187 individuals). Red colour corresponds to  indifferent chicks and blue colour corresponds to imprinted chicks. The bigger values correspond to higher level of imprinting}
\label{Fig13}
\end{figure}

A quantitative analysis of the imprinting level of imprinted and indifferent chicks (187 individuals) was performed (as described in Sec. \ref{sec:classificationRes}), with results shown in Fig.~\ref{Fig13}. It shows that the majority of indifferent chicks have an imprinting index close to $-20$, with small imprinting variability: it can be explained by the fact that indifferent chicks are immobile during most of the experiments. Imprinted chicks have a larger imprinting variability than indifferent chicks. Most imprinted chicks have an imprinting index between $0$ and $65$.
\\

\subsection{Time evolution of social bonding}
\begin{figure}[]
\centering\includegraphics[width=8.3cm]{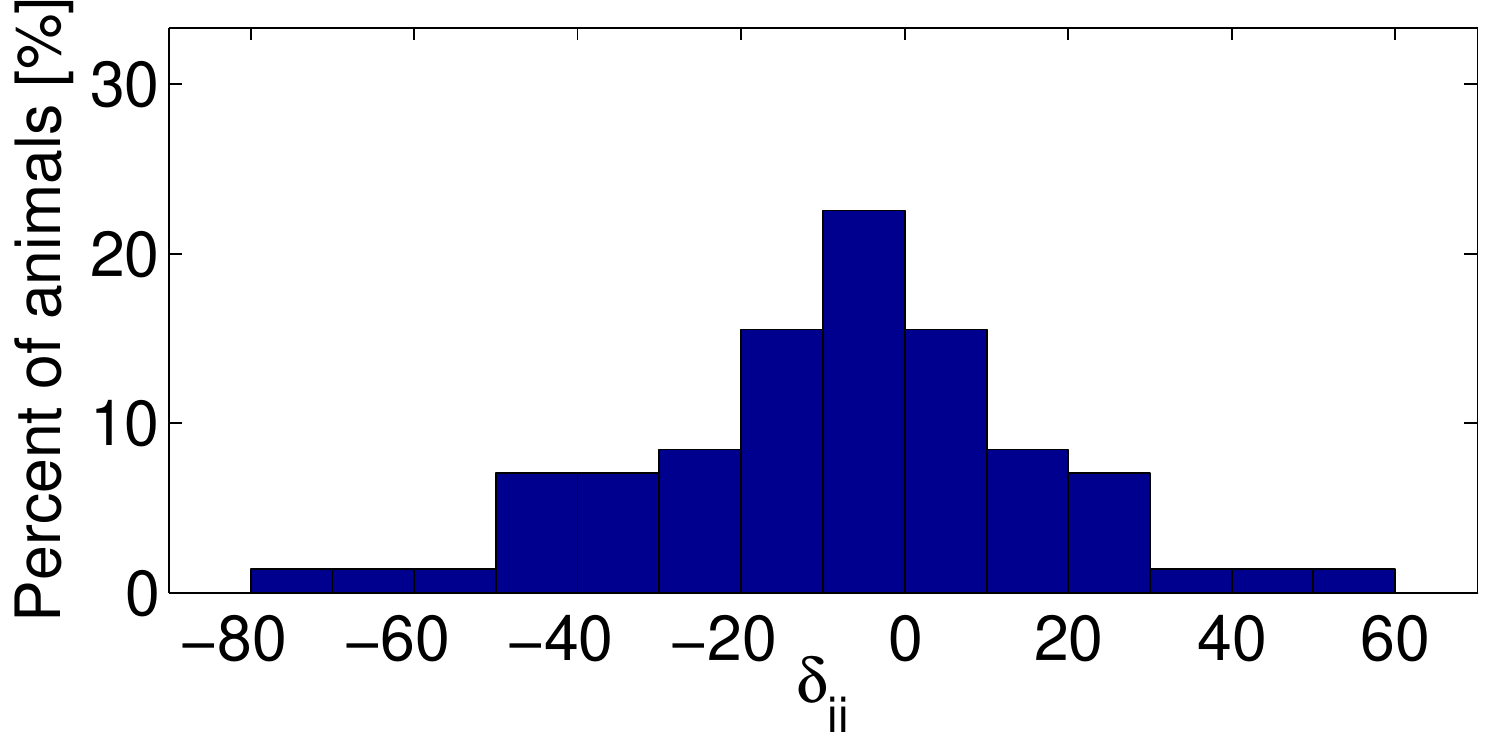}
\caption{{\bf Imprinting indexes evolution distribution.} We show here a distribution of difference in the imprinting index between the second and the first individual tests for 71 individuals. Mean value of this dataset is $-7.5018$, thus we observe a general decline in the individual attachment towards the robot, possibly due to the aging of the animals. Note that this distribution is not normal, even if it looks the part (the Kolmogorov-Smirnoff test fails when compared to the normal distribution, with $p=0.4163$).}
\label{Fig14}
\end{figure}

We investigate the influence of social bonding on the imprinting quality of chicks toward the robot. After the imprinting process, $72$ randomly selected chicks are put in $12$ groups of $6$ individuals for $20$ days. Each of these group contains a random number of imprinted, indifferent and avoiders chicks. Then, we assess if the chicks have changed their individual behaviour toward the robot by performing another set of experiments, with the same approach described in previous sections. These experiments were only performed on $42$ individuals in $7$ groups.
We produce imprinting indexes using the method described in \ref{sec:indivVar}.
To study the change in imprinting quality of chicks put in groups compared to chicks alone, we introduce the measure:
$$
\delta_{ii} = ii_{t_2} - ii_{t_1}
$$
where $ii_{t_1}$ and $ii_{t_2}$ are the imprinting index of a chick respectively before and after being put into groups.
The distribution of the $\delta_{ii}$ measure is shown in Fig.~\ref{Fig14}. The mean value of all considered $\delta_{ii}$ is $-7.5018$: we observe a general decline of individual attachment toward the robot. It can possibly be explained by the aging of the animals.
\\

\begin{figure}[]
\centering\includegraphics[width=8.3cm]{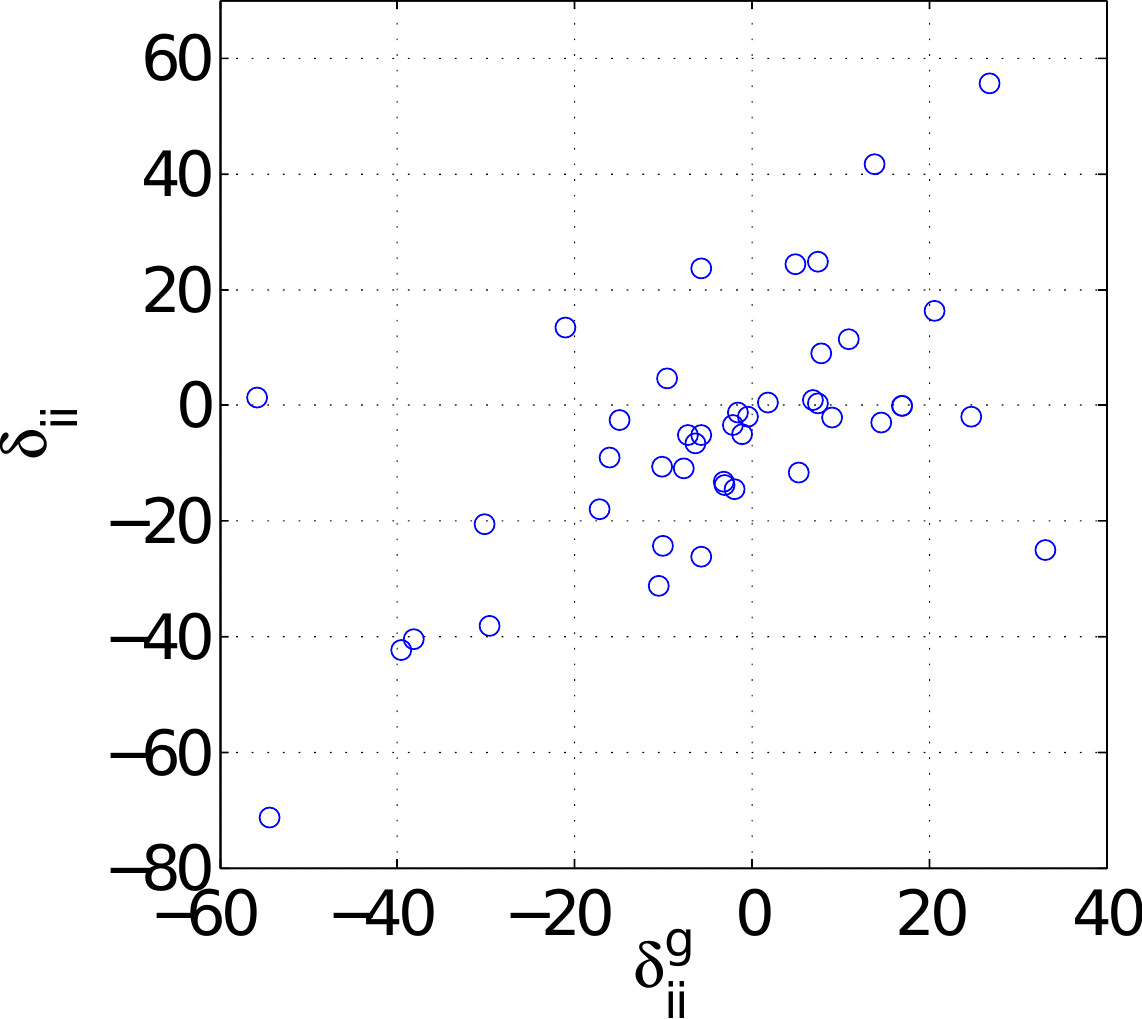}
\caption{{\bf Quantifying social effects on the temporal change of imprinting.} Link between imprinting strength evolution and deviation of the chick from its group based on the change of the individual imprinting index after experiments of $42$ chicks in $7$ groups of $6$ chicks presenting various level of imprinting. After being put in groups, chicks have an imprinting quality toward the robot that is closer to the average of their group. A chick that originally was not very interested in the robot develops a stronger attachment after being placed in the group of individuals strongly imprinted toward the robot. A chick that originally was very attached to the robot become less interested in the robot after being placed in the group of individuals that have a low interest toward the robot.}
\label{Fig15}
\end{figure}

To take into account the composition of the groups in our analysis, we introduce the \textit{group imprinting index} as an average of imprinting indices of group members in the first individual test:
$ii^g = \sum_{k=1}^6 ii^{k}_{t_1}/6$ where $ii^{k}_{t_1}$ is the imprinting index of the $k$-th chick in a group before being put into a group.
We compute $\delta_{ii}^{g}$ as a difference between the group average and the individual imprinting index: $\delta_{ii}^{g} = ii^g-ii$. Chicks for which this value is positive shows an imprinting quality below the average in their group, and chicks for which this value is negative show more attachment toward the robot compared to the average of their group.
The link between the $\delta_{ii}^g$ and $\delta_{ii}$ is shown in Fig.~\ref{Fig15}. The represented distributions are significantly positively correlated (Student's T test for a transformation of the Pearson's correlation $p=0.000012$, with a correlation coefficient of $0.5907$).
We observe that, after being put in groups, chicks have an imprinting quality toward the robot that is closer to the average of their group. A chick that originally was not very interested in the robot develops a stronger attachment after being placed in the group of individuals strongly imprinted toward the robot. On the other hand, a chick that originally was very attached to the robot become less interested in the robot after being placed in the group of individuals that have a low interest toward the robot.
\\

%%%%%%%%%%%%%%%%%%%%%%%%%%%%%%%%%%%%%%%%%%%%%%%%%%%%%%%%%%%%%%%%%%%%%%%%%%%%%%%%%%%%%%%%%%%%%%%%%%%%%%%%%
%%%%%%%%%%%%%%%%%%%%%%%%%%%%%%%%%%%%%%%%%%%%%%%%%%%%%%%%%%%%%%%%%%%%%%%%%%%%%%%%%%%%%%%%%%%%%%%%%%%%%%%%%
%%%%%%%%%%%%%%%%%%%%%%%%%%%%%%%%%%%%%%%%%%%%%%%%%%%%%%%%%%%%%%%%%%%%%%%%%%%%%%%%%%%%%%%%%%%%%%%%%%%%%%%%%
%%%%%%%%%%%%%%%%%%%%%%%%%%%%%%%%%%%%%%%%% MATERIALS AND METHODS %%%%%%%%%%%%%%%%%%%%%%%%%%%%%%%%%%%%%%%%%
%%%%%%%%%%%%%%%%%%%%%%%%%%%%%%%%%%%%%%%%%%%%%%%%%%%%%%%%%%%%%%%%%%%%%%%%%%%%%%%%%%%%%%%%%%%%%%%%%%%%%%%%%
%%%%%%%%%%%%%%%%%%%%%%%%%%%%%%%%%%%%%%%%%%%%%%%%%%%%%%%%%%%%%%%%%%%%%%%%%%%%%%%%%%%%%%%%%%%%%%%%%%%%%%%%%
%%%%%%%%%%%%%%%%%%%%%%%%%%%%%%%%%%%%%%%%%%%%%%%%%%%%%%%%%%%%%%%%%%%%%%%%%%%%%%%%%%%%%%%%%%%%%%%%%%%%%%%%%

% You may title this section "Methods" or "Models".
% "Models" is not a valid title for PLoS ONE authors. However, PLoS ONE
% authors may use "Analysis"
\section{Materials and Methods} \label{sec:methods}

{\bf Ethics statement} Animal experiments were performed in accordance with the recommendations and guidelines of the local competent authority and ethic committee.

\subsection{The PoulBot robot} \label{sec:poulbot}
The robot does not look like a chicken, however chicks developed a strong social attachment to it thanks to the filial imprinting mechanism \cite{Rogers1996}. Chicks prefer imprinting object with spotted pattern, dots or strikes \cite{Klopfer1964a}. Moving objects are more attractive \cite{Bateson1966}, as visual imprinting would occur on the association between object form and movement \cite{Broom1969}. Chicks are attracted by colours that are easy to discriminate from the colours of the ground. Red-orange or blue revealed to be the most attractive colours \cite{Ham2007}. Not only the nature but the characteristics of the colours have an influence on recognition and imprinting \cite{Davis1976}. Chick can learn or be imprinted on static coloured two-dimensional stimuli \cite{Salzena1971}. Contrasted patterns attract more the chick attention and are more effective for accurate memory \cite{Osorio1999}.
\\

Auditory communication is of importance in poultry \cite{Mench2001} and imprinting also depends on auditory stimuli \cite{VanKampen1991,Mench2001}. Early exposure to auditory stimulus induces preferences in experiments. Several vocalizations and their roles still need to be investigated in imprinting process but it is known that early exposure to a sound influence the later preferences. It seems that both auditory and visual learning are enhanced during the critical period, which is consistent with the interpretation that imprinting is a within-event learning occurring when elements and their representations are linked \cite{Bolhuis1999}.
\\

The main features of the robot namely movement, colour pattern and sound emission have been designed taking into account these results. Moreover, each of them can be switched on and off and programmed independently of each other allowing the study of the multi-modality of factors favoring filial imprinting. The fact that the robot is fully autonomous and can move in the same environment as the chicks allows studies where the animals can move freely and in
groups, thus allowing to study the link between individual and collective behaviour.
\\

The mobile robot used in this study is a track-type mobile robot presented in Fig.~\ref{Fig1} \cite{Gribovskiy2010}. We named it the \textit{PoulBot}. The robot is a configuration of the marXbot robot -- a modular research robot developed at EPFL \cite{Bonani2010}. The PoulBot consists of the following modules: a \textit{base module} providing energy, means for locomotion, a gyroscope, an accelerometer, 24 short range infrared proximity sensors around the robot and a speaker, a \textit{colour pattern module} allowing a robot to carry a specific pattern, an extra \textit{ring of proximity sensors} placed above the colour pattern module and a \textit{main computer board} with a 533MHz Freescale i.MX31 and an omnidirectional camera, the main board also provides Wi-Fi and Bluetooth wireless connectivity. On the top of the robot we fixed three colour markers used to track the robot position and orientation with a camera fixed above the experimental setup (see below). Every module of the robot consists of one or several
Microchip dsPIC33 microcontrollers that drive the sensors and/or actuators of the module. The microcontrollers are connected with each other and with the main computer of the robot through a controller-area network (CAN) bus. The robot can be extended with the sound acquisition module and with the pecking module.
\\

The control system of our robot is a hierarchical behaviour based controller \cite{Mataric2008}. The robot is equipped with a set of primitive behaviours tightly bonded with the sensors and actuators of the robot. Each primitive behaviour serves to achieve a particular goal or to perform a specific activity (e.g., \textit{wall-following} or \textit{obstacle-avoidance} behaviours). The primitive behaviours are combined together to form higher level composite behaviours for specific experiments. Behaviour-based systems work well in dynamic environments, in cases when fast reaction and high adaptability are important. These characteristics make the behaviour based approach a natural choice when designing a control system for a robot that interacts with animals.

\subsection{Animal hatching housing and handling}
We performed experiments with chicks, \textit{Gallus gallus domesticus}, from hatching up to $3$ week-old. As breed, we chose the egg layer White Leghorn that is common in scientific studies. The incubators were disposed together in a $10 m^2$ isolated rearing room that was maintained in the dark and at a constant temperature of $35^{\circ}C$ during incubation. Eggs were incubated at the classical temperature of $38^{\circ} C$ with a relative humidity of $70\%$. After hatching, chicks were identified with numbered plastic rings their hatching time was recorded. They were left to rest and dry in the incubators. When dry, they were kept in the dark in individual boxes until the imprinting process. The light was left off to reduce imprinting of first hatched chicks on siblings. The presence of humans was reduced to the minimum necessary for handling. During hatching, the frequency of visit by experimentalists was of $1$ hour.
\\

During incubation and up to the imprinting process, two speakers and a MP3 device played a calibrated sound sound that the robots also played during imprinting and tests. The sound was played from egg state to the end of the imprinting sessions. This calibrated sound had a frequency of $6 kHz$ and with a beep duration of $150 ms$ and interval between two beeps of $350 ms$. These features have been chosen in correspondence to the range of chick audition \cite{Collias1953,Collias1987}. We chose the upper threshold of the range and a sound designed to be neutral since it is totally artificial, which means that it is not related to a natural chick or hen calls, such as alarm calls or clucking, and have thus no biological sense. This was made for auditory imprinting not oriented towards a special message signal.
\\

After the imprinting process, chicks are kept in groups of $6$ in breeding nurseries and given food and water \textit{ad libitum}. The temperature of the breeding room was reduced during the three weeks from $32$ to $23^{\circ}C$ according to breeding standards.

\subsection{Experimental setup and protocol}
\begin{figure*}[]
\centering\includegraphics[width=14.00cm]{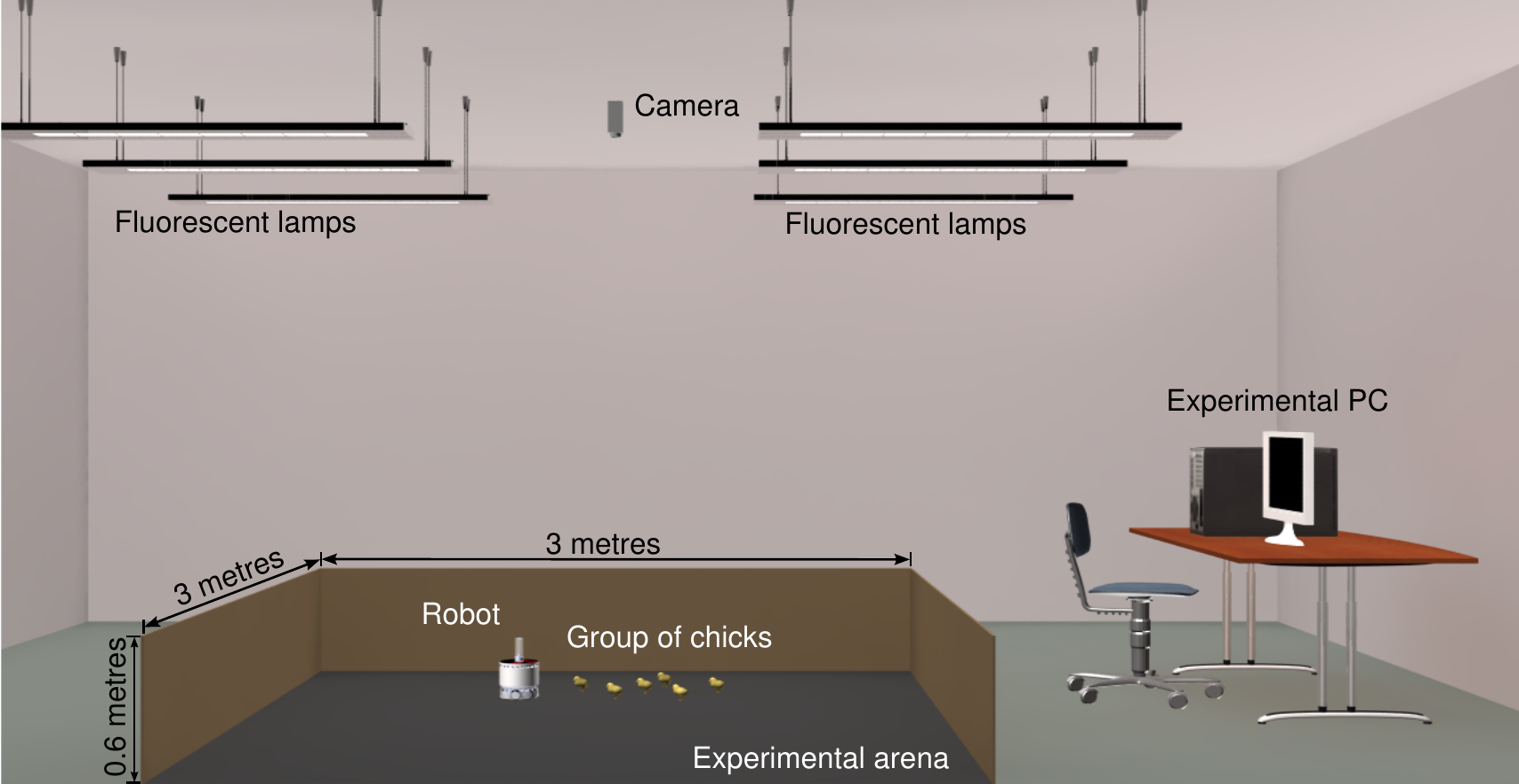}
\caption{{\bf Experimental facilities.} The experimental arena is a square of $3$~m side length surrounded by walls made of $0.6$~m high wood boards. The arena is filmed by a digital camera at 10 frames/s connected to a computer by a gigabyte Ethernet link. During the experiments there is no human in the experimental room as the system is fully autonomous. The computer records and processes the digital video of a trial in real time. It can also monitor and display a trial remotely for human supervision. The computer can also monitor the robot and send instruction to it, if necessary. Animals and robots are introduced in the set-up and let free to move around in the arena for 30 minutes by trials.}
\label{Fig16}
\end{figure*}

\begin{figure*}[]
\centering\includegraphics[width=14.00cm]{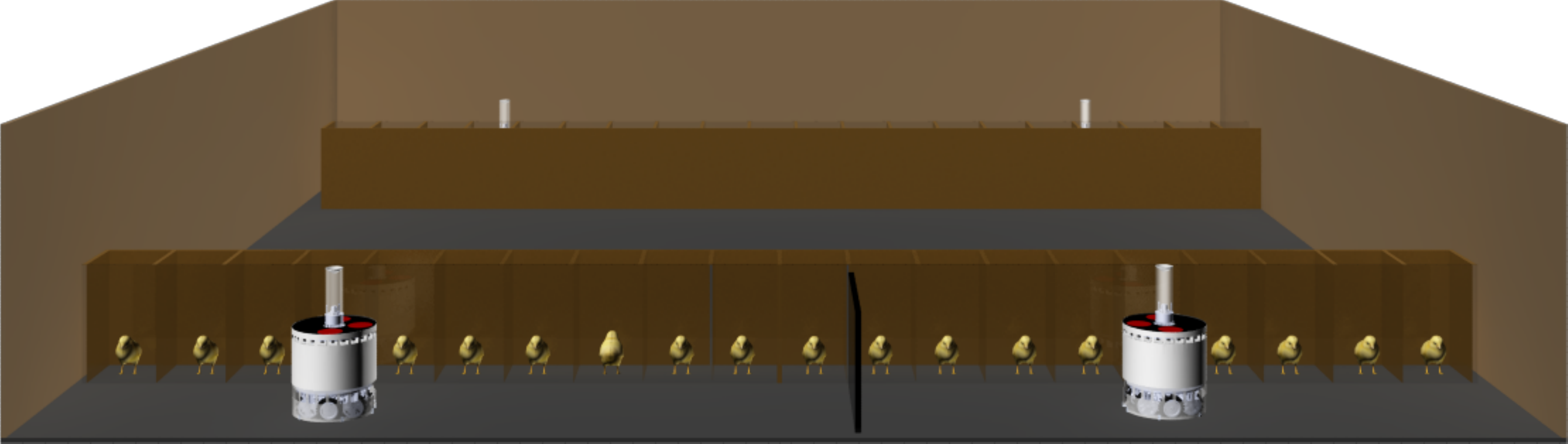}
\caption{{\bf Design of the imprinting set-up}. The robots travel back and forth between two walls in front of the chicks.}
\label{Fig17}
\end{figure*}

We ran open-field tests, where animals and robots were released on a flat arena surrounded by walls and their behaviour was studied. The experimental arena is a flat square  of $3$~m by $3$~m, surrounded by a wooden wall of $60$ cm in height. A common daylight lamp is a source of a strong infrared (IR) emission that affects the IR sensors of the robot. To resolve this issue, we used lamps with reduced infrared emission FQ49W/965 by OSRAM. Twelve lamps were uniformly fixed on the ceiling to provide lighting conditions as homogeneous as possible. A Scout scA1000-30gc colour camera by Basler Vision Technologies with a CS-mount T3Z2910CS varifocal lens by Computar was fixed above the setup for tracking and recording tasks. The camera has a resolution of $1032\times778$ pixels and a maximum frame rate of $30$ frames per second. It uses a Gigabit Ethernet connection to transfer video to the PC. The experimental PC has an eight-core Intel Core$^{\rm TM}$2 Quad  processor and 2 GB of RAM. It runs the monitoring and
recording software presented in the next Section  and the GUI module of the robot control system. It also serves as a temporary storage of the experimental data recorded during the day, before the data is transferred to external hard drives. The temperature in the experimental room during the whole experimentation period was kept constant. The experimental facilities are described in Fig.~\ref{Fig16}. The imprinting set-up is shown in Fig.~\ref{Fig17}.
\\

The imprinting procedure precedes every series of experiments. Depending on birth rates, we tried to compose imprinting batches with chicks born within at most two hour intervals. This was made to have a sufficient number of chicks per imprinting session. To perform the imprinting we build an arena composed of two rows of boxes on the sides. Four boxes were built in MDF wood with a Plexiglas side facing the wall to let chicks see the robot that traveled back and forth in front of them. Wooden boxes were $1500$x$150$x$300$ mm. Each box had ten compartments of $150$x$150$ mm to house one chick at a time. By putting one chick per box the setup allowed the simultaneous imprinting of 40 chicks. Four robots moved along the walls in front of each box at a close distance of about $100$ mm , forward and backward, at a speed of $60$ mm/s, one in front of each box. The robots emitted a calibrate sounds that was the same as during incubation and hatching (see above) and displayed a specific colour pattern. This individual
and in-line imprinting process was designed to reduce imprinting among siblings. We performed the imprinting sessions according to the optimal period for chicken filial imprinting the The first exposure to the robots was 9 hours after hatching \cite{Ramsay1954,Hess1973}. The first exposure was programmed 9 hours after the birth of the chicks. Three sessions of one hour in the presence of the robot were done interrupted in-between by one hour resting time in the breeding room. Rotation were done to imprint all chicks of a hatching batch (100 individuals at most).
\\

Shortly after an end of the imprinting procedure we tested all chicks: every chick was left for a half an hour with a robot \textit{wandering} on the experimental arena and chick behaviour was observed. The robot emits a calibrated beeping sound and moves straight at a constant speed of $70$ mm/s. When the robot reaches an edge of the set-up it bounces back with a
random angle. The speed of the robot was selected according to observation of the mean chick speed during previous calibration experiments. Each test lasted for 30 minutes and for data
analysis the first 2 minutes are skipped and the next 25 minutes are kept. This avoids artifacts due to animal introduction and extraction from the experimental setup. All tests were
recorded and stored as digital videos for further processing. The chicks were tested, one after the other, in the sequence corresponding to their age, older first, younger last.

\subsection{Classification of filial imprinting} \label{sec:classificationMethod}
For classifying chicks imprinting success, as described in Sec. \ref{sec:classificationRes}, we used a linear support vector machine (SVM) classifier, a supervised learning method that is often used to solve problems in classification, regression, and novelty detection \cite{Vapnik1995,Meyer2003,Bishop2006}.
\\

Classification using linear SVM is performed by solving the problem of finding a hyperplane partitioning the training dataset. During training, the examples of separate classes are separated by a gap that is as wide as possible. We make the hypothesis that our exploration space can be linearly separated, so using linear SVM (instead of the non-linear SVM using kernel function \cite{Shawe-Taylor2004}) is sufficient. SVM is fundamentally a two-class classifier, but a number of methods exist to handle a higher number of classes \cite{Bishop2006}.
\\

The LIBSVM (Version 3.0) \cite{Chang2001} was used to perform the classification.
\\

We used the following statistics on chicks and robot movements as relevant features (explanatory variables) for the classification process: \textit{mean distance between a robot and a chick}, \textit{mean speed of a chick}, and its \textit{standard deviation}.
We reduced the dimensionality of the related feature vector from three to two by using principal components analysis (PCA) \cite{Bishop2006}. The resulting dataset was used to train a three-class SVM classifier.
\\

The training takes about 6~milliseconds (with 64 individuals) and the classification takes about 7~milliseconds (for 205 individuals) with Matlab R2011b running on an ordinary laptop PC (Lenovo Thinkpad T60p, Linux Ubuntu 12.04). Moreover, for individual tests the extraction of the chick and robot positions is done automatically in real time during the recording of the test. Thus the classifier allows to categorize, in real time, rapidly, large batches of experiments saving a huge amount of experimental time.

\subsection{Segmentation of trajectories}\label{sec:behaviouralMeasurments}
The recorded experimental videos were analyzed during the experimental run using a SwisTrack that is an open source multi-platform tracking software \cite{Lochmatter2008}. The robot position and orientation are defined by three colour markers on top of it. Thus the detection of robots and chicks is colour based. The coordinates of the robots and the animals extracted from the video were mapped to the real-world coordinates (in mm) by using the calibration routine based on the well known Tsai's calibration technique \cite{Tsai1987}. In order to remove high frequency noise introduced into trajectories by the detection errors, every trajectory is filtered by using the Savitzky-Golay smoothing filter \cite{Savitzky1964}.
\\

This trajectories are then segmented. The segmentation process identify segments of the trajectories (i.e.: group of continuous time-steps in the chicks behavioural traces) when the chick exhibit a particular behaviour. This is a first step toward labelling the trajectories with the observed behaviour of the chicks toward the robot. In an optimal segmentation, segments should end when a chick changes into behaviour.
A typical segmentation method used in the literature is the $N$-steps window segmentation: trajectories containing $M$ time-steps are split into $M/N$ segments each with a size of $N$ time-steps. An extreme case is the $1$-step window segmentation, where every time-step is a new segment. Note that the choice of $N$ can influence the reliability of a segmentation: if $N$ is too big, segments can include period of times where the chicks exhibit several different behaviours. On the other hand, $N$-step window segmentation with a low $N$ can be more sensitive to noise.
We introduce a more robust segmentation method, the 'Threshold crossing' method.
We make two hypothesises, supported by human observations: first, that chicks reduce their speed below a given threshold when they change behaviour; second, that chicks accelerate when they begin to exhibit a given behaviour, and decelerate when they change behaviour. This allow us to introduce an original segmentation method, that we call 'Threshold crossing'. This segmentation method is based on the speed and acceleration of a chick (Fig. \ref{Fig7}): time-steps when the speed or acceleration of a chicks are below predefined thresholds are considered as stops.

\subsection{Classification and Clustering of trajectories} \label{sec:classificationAndClusteringOfTrajectories}
To find the quantitative representation of these behaviours in the parameter space we use either classification algorithms or clustering algorithms.
To prepare the training set we manually labeled a number of segments of trajectories, obtained using the segmentation method described in \ref{sec:behaviouralMeasurments}. While classification algorithms are trained on human-labelled data, clustering algorithms identify behaviours without any training. Either techniques will label observation data into 7 classes of behaviours: \textit{"stops", "joins", "follows", "in front", "loops","bumped"} and \textit{"other"}, which reflect the activity of the chick, and its relation to the movements of the robot. The Python library scikit-learn was used to perform all classification and clustering operations.\\

(1) Decision Trees (\cite{Breiman1984}, \textbf{DT} in Table~\ref{table1}).
Decision Trees correspond to a recursive partitioning of the training set into several homogeneous subsets. It has a flowchart structure where nodes and branches represent conditions (expressing how to partition the training set), and leafs nodes represent class labels.
Decision Trees are non-parameter and nonlinear, and so can be used in cases when little is known a-priori about the relationship between the variables. They are also simple to interpret and verify by a human.
The classification could incur a loss due to overfitting, a pruning mechanism \cite{Helmbold1995} was used to solve this issue.\\

(2) Random Forests (\textbf{DT/Forest} in Table~\ref{table1}). It is a variation of the Decision Tree method: one can use an ensemble of trees, that is known as a random forest \cite{Breiman2001}, instead of a single tree, to increase the accuracy of classification.\\

(3) A k-Nearest Neighbours classifier, with an euclidean distance metric \cite{kantardzic2011data}. This algorithm makes use of a measure of similarity between all the training records and the features of the object to be classified, then identify the most similar $k$ neighbours of all objects, and determine each object class by an object by a majority vote of its neighbours (i.e. assign the class which is the most frequent among the $k$ training records nearest to that object). Note that kNN classifiers are known to be sensitives to noisy, irrelevant features, or by the difference of scaling of the features \cite{kantardzic2011data}. That may explain the classification results (Table~\ref{table1}), where kNN was out-performed by other algorithms.\\

(4) Linear multi-class SVM classifier, as described in Section \ref{sec:classificationMethod}.\\

(5) K-Means \cite{kantardzic2011data}. This clustering algorithm partitions the observations into $K=7$ clusters, in which each observation belongs to the cluster with the nearest mean. The results are akin to partitioning the data space into Voronoi cells. Convergence speed can be greatly affected by the choice of initial values. We use the popular k-means++ \cite{arthur2007k} algorithm for choosing initial values.\\

%%%%%%%%%%%%%%%%%%%%%%%%%%%%%%%%%%%%%%%%%%%%%%%%%%%%%%%%%%%%%%%%%%%%%%%%%%%%%%%%%%%%%%%%%%%%%%%%%%%%%%%%%
%%%%%%%%%%%%%%%%%%%%%%%%%%%%%%%%%%%%%%%%%%%%%%%%%%%%%%%%%%%%%%%%%%%%%%%%%%%%%%%%%%%%%%%%%%%%%%%%%%%%%%%%%
%%%%%%%%%%%%%%%%%%%%%%%%%%%%%%%%%%%%%%%%%%%%%%%%%%%%%%%%%%%%%%%%%%%%%%%%%%%%%%%%%%%%%%%%%%%%%%%%%%%%%%%%%
%%%%%%%%%%%%%%%%%%%%%%%%%%%%%%%%%%%%%%%%%%%%%% DISCUSSION %%%%%%%%%%%%%%%%%%%%%%%%%%%%%%%%%%%%%%%%%%%%%%%
%%%%%%%%%%%%%%%%%%%%%%%%%%%%%%%%%%%%%%%%%%%%%%%%%%%%%%%%%%%%%%%%%%%%%%%%%%%%%%%%%%%%%%%%%%%%%%%%%%%%%%%%%
%%%%%%%%%%%%%%%%%%%%%%%%%%%%%%%%%%%%%%%%%%%%%%%%%%%%%%%%%%%%%%%%%%%%%%%%%%%%%%%%%%%%%%%%%%%%%%%%%%%%%%%%%
%%%%%%%%%%%%%%%%%%%%%%%%%%%%%%%%%%%%%%%%%%%%%%%%%%%%%%%%%%%%%%%%%%%%%%%%%%%%%%%%%%%%%%%%%%%%%%%%%%%%%%%%%
\section{Discussion}
This study validates the use of imprinting to ensure a social bond between robots and animals in order to perform controlled experiments with calibrated stimuli and sustained social interactions.
When the robots are accepted by the animals, they can be programmed to keep sustained social interactions with them in long duration trials (our experiments lasted 30 minutes).
As the robots are autonomous and work in close loop of interactions with the animals,they allow to avoid any human interference during the trials. The whole set-up presented here works in absence of humans in the experimental room and without supervision. The robots can be programmed to produce calibrated and repetitive multi-modal stimuli including visual (colour patterns), sound (prerecorded natural cues or totally artificial) and motion patterns.
\\

Moreover, we build a framework that allows automation of high throughput data analysis. From the positions of animals and their dynamics in time we construct quantitative automated ethograms.
\\

We test state-of-the-art algorithms to automate data analysis and show that supervised method based on Decision Tree with forest gives the best results. We also show that simple unsupervised methods such as K-means already gives interesting results. Here, the algorithms allow to combine the knowledge of human experts with fast computational methods. It is then possible to analyze the experimental data of 205 animals in a few minutes on modern computers.
\\

This paper has two contributions.
First, we introduce a methodology that allows the automated generation of quantified ethograms both on the individual level (represented by sequences of labelled behaviours) and on the group level (represented by Finite State Machines).
Second, we investigate the individual variability of chicks in term of imprinting, using a classification algorithm.
Results also confirm that social bonding have an influence on the imprinting quality. We observe that, after being put in groups, chicks have an imprinting quality toward the robot that is closer to the average of their group. A chick that originally was not very interested in the robot develops a stronger attachment after being placed in the group of individuals strongly imprinted toward the robot. On the other hand, a chick that originally was very attached to the robot become less interested in the robot after being placed in the group of individuals that have a low interest toward the robot.
\\

This methodology is a first step towards mathematical modelling based on quantified ethograms, and this study is a first step toward making the link between inter-individual variability and collective behaviour.
In our methodology, we first study and quantify individually each chick. Then, the chicks are put in different grouping configuration, according to their inter-individual variability.
An interesting and on-going perspective is to make a group behaviour analysis taking in account the chicks individual behavioural traits, and the group composition.

\section*{Acknowledgment}
This work is supported by the Swiss National Science Foundation grant no. 112150 "Mixed society of robots and vertebrates",  EU-ICT project "ASSISI-bf", no. 601074.

\end{document}